\documentclass[journal]{IEEEtran}
\ifCLASSINFOpdf
\else
\usepackage[dvips]{graphicx}
\fi
\usepackage{hyperref}
\hypersetup{
    colorlinks=true,
    linkcolor={black},
    citecolor={black},
    urlcolor={black}
    }
\usepackage{cite}
\usepackage{tumcolor}
\usepackage{amsmath,amssymb,amsthm,fixmath}
\usepackage{mathtools}
\usepackage{algorithm}
\usepackage{algorithmic}
\usepackage{xcolor}
\usepackage{bm}
\usepackage{makecell}
\usepackage{pgfplots}
\pgfplotsset{compat=1.17}
\usepackage{tikz}
\usetikzlibrary{intersections,pgfplots.fillbetween,patterns,spy,shapes.geometric,calc,positioning}
\usetikzlibrary{arrows,decorations.pathmorphing,backgrounds,fit,matrix,fillbetween}
\usetikzlibrary{shapes.geometric, arrows.meta}
\usetikzlibrary{decorations.pathreplacing,decorations.markings}
\usepgfplotslibrary{groupplots}
\usepackage{subcaption}
\usepackage{tabularx}
\usepackage{soul}
\usepackage[acronym,shortcuts]{glossaries}
\usepackage{cleveref}
\definecolor{mylila}{RGB}{153,50,204}

\newcommand{\op}[1]{{\operatorname{#1}}}
\newcommand{\B}[1]{{\bm{#1}}}
\newcommand{\uproman}[1]{\uppercase\expandafter{\romannumeral#1}}

\newcommand{\sign}{\operatorname{sign}}

\newcommand{\dx}{\operatorname{d}\hspace{-1.5pt}}
\newcommand{\h}{^{\operatorname{H}}}
\newcommand{\T}{^{\operatorname{T}}}
\newcommand{\inv}{^{-1}}
\newcommand{\diag}{\operatorname{diag}}
\newcommand{\nondiag}{\operatorname{nondiag}}
\DeclareMathOperator{\eye}{\mathbf{I}}
\DeclareMathOperator{\dft}{\mathbf{F}}
\newcommand{\vect}{\operatorname{vec}}
\newcommand{\tr}{\operatorname{tr}}
\newcommand{\sqrtm}{^{-\frac{1}{2}}}
\newcommand{\C}{\mathbb{C}}

\newcommand{\erf}{\operatorname{erf}}

\newcommand{\NC}{\mathcal{N}_{\C}}
\newcommand{\N}{\mathcal{N}}
\newcommand{\DKL}{\op D_{\text{KL}}}

\newacronym{mimo}{MIMO}{multiple-input multiple-output}
\newacronym{simo}{SIMO}{single-input multiple-output}
\newacronym{mse}{MSE}{mean square error}
\newacronym{cme}{CME}{conditional mean estimator}
\newacronym{pdf}{PDF}{probability density function}
\newacronym{adc}{ADC}{analog-to-digital converter}
\newacronym{mmse}{MMSE}{Minimum mean square error}
\newacronym{snr}{SNR}{signal-to-noise ratio}
\newacronym{evd}{EVD}{eigenvalue decomposition}
\newacronym{crb}{CRB}{Cram\'er-Rao bound}
\newacronym{map}{MAP}{maximum a posteriori}
\newacronym{cs}{CS}{compressive sensing}
\newacronym{ls}{LS}{least squares}
\newacronym{awgn}{AWGN}{additive white Gaussian noise}
\newacronym{csi}{CSI}{channel state information}
\newacronym{ml}{ML}{maximum likelihood}
\newacronym{pmf}{PMF}{probability mass function}
\newacronym{cdf}{CDF}{cumulative distribution function}
\newacronym{rv}{RV}{random variable}
\newacronym{gmm}{GMM}{Gaussian mixture model}
\newacronym{mt}{MT}{mobile terminal}
\newacronym{bs}{BS}{base station}
\newacronym{acg}{AGC}{automatic gain control}
\newacronym{ula}{ULA}{uniform linear array}
\newacronym{upa}{UPA}{uniform planar array}
\newacronym{mfa}{MFA}{mixture of factor analyzer}
\newacronym{vae}{VAE}{variational autoencoder}
\newacronym{em}{EM}{expectation-maximization}
\newacronym{dft}{DFT}{discrete Fourier transform}
\newacronym{elbo}{ELBO}{evidence lower bound}
\newacronym{mmwave}{mmWave}{millimeter wave}
\newacronym{rf}{RF}{radio frequency}
\newacronym{gamp}{GAMP}{generalized approximate message passing}
\newacronym{iht}{IHT}{iterative hard thresholding}
\newacronym{fft}{FFT}{fast Fourier transform}
\newacronym{psd}{PSD}{positive semi-definite}
\newacronym{nn}{NN}{neural network}
\newacronym{relu}{ReLU}{rectified linear unit}
\newacronym{uatf}{UF}{use-and-then-forget}
\newacronym{mrc}{MRC}{maximum-ratio combining}
\newacronym{los}{LOS}{line-of-sight}
\newacronym{nlos}{NLOS}{non-line-of-sight}
\newacronym{3gpp}{3GPP}{3rd Generation Partnership Project}
\newacronym{cnn}{CNN}{convolutional \ac{nn}}
\newacronym{aqnm}{AQNM}{additive quantization noise model}
\newacronym{vi}{VI}{variational inference}

\newcolumntype{Y}{>{\centering\arraybackslash}X}

\colorlet{TUMOrange80}{TUMOrange!80}%
\colorlet{TUMOrange60}{TUMOrange!60}%
\colorlet{TUMOrange50}{TUMOrange!50}%
\colorlet{TUMOrange40}{TUMOrange!40}%
\colorlet{TUMOrange20}{TUMOrange!20}%

\colorlet{TUMGreen80}{TUMBeamerGreen!80}%
\colorlet{TUMGreen60}{TUMBeamerGreen!60}%
\colorlet{TUMGreen50}{TUMBeamerGreen!50}%
\colorlet{TUMGreen40}{TUMBeamerGreen!40}%
\colorlet{TUMGreen20}{TUMBeamerGreen!20}%

\colorlet{TUMRed80}{TUMBeamerDarkRed!80}%
\colorlet{TUMRed60}{TUMBeamerDarkRed!60}%
\colorlet{TUMRed50}{TUMBeamerDarkRed!50}%
\colorlet{TUMRed40}{TUMBeamerDarkRed!40}%
\colorlet{TUMRed20}{TUMBeamerDarkRed!20}%

\tikzset{genie/.style={mark options={solid},color=TUMBeamerRed, line width=\lineWidth}}
\tikzset{global/.style={mark options={solid},color=TUMBeamerOrange, line width=\lineWidth, mark=o, mark size=\marksize}}
\tikzset{gmm/.style={mark options={solid},color=TUMBeamerBlue, line width=\lineWidth, mark=square, mark size=\marksize, dashed}}
\tikzset{LS/.style={mark options={solid},color=gray, line width=\lineWidth, mark=triangle, mark size=\marksize}}
\tikzset{gmm_toep/.style={mark options={solid},color=TUMBeamerLightBlue, line width=\lineWidth, mark=diamond, mark size=\marksize, dashdotted}}
\tikzset{gmm_circ/.style={mark options={solid},color=TUMBlueDarker, line width=\lineWidth, mark=x, mark size=\marksize, dashdotted}}
\tikzset{em_gm_gamp/.style={mark options={solid},color=TUMBeamerGreen, line width=\lineWidth, mark=pentagon, mark size=\marksize}}
\tikzset{vae_noisy/.style={mark options={solid},color=mylila, line width=\lineWidth, mark=o, mark size=\marksize, dashed}}
\tikzset{mfa/.style={mark options={solid},color=brown, line width=\lineWidth, mark=pentagon, mark size=\marksize, dashed}}
\tikzset{deep/.style={mark options={solid},color=black, line width=\lineWidth, mark=|, mark size=\marksize}}
\tikzset{vae_quant/.style={mark options={solid},color=TUMBeamerRed, line width=\lineWidth, mark=diamond, mark size=\marksize}}
\tikzset{gmm_quant/.style={mark options={solid},color=TUMBeamerGreen, line width=\lineWidth, mark=triangle, mark size=\marksize}}

\newcommand{\lineWidth}{1.0pt}
\newcommand{\marksize}{1.6pt}

\pgfplotsset{
  every axis legend/.append style={
    font=\scriptsize 
  }
}

\pgfplotsset{
  every axis title/.append style={
    font=\scriptsize 
  }
}

\pgfplotsset{every axis/.append style={
                    label style={font=\scriptsize},
                    tick label style={font=\scriptsize}  
                    }}

\newtheorem{lemma}{Lemma}

\glsdisablehyper

\begin{document}
	\bstctlcite{IEEEexample:BSTcontrol}

\title{Channel Estimation for Quantized Systems based on Conditionally Gaussian Latent Models}

\author{Benedikt Fesl, \IEEEmembership{Graduate Student Member, IEEE}, Nurettin Turan, \IEEEmembership{Graduate Student Member, IEEE}, \\ Benedikt B\"ock, \IEEEmembership{Graduate Student Member, IEEE}, and Wolfgang Utschick, \IEEEmembership{Fellow, IEEE}
	\thanks{
	}
	\thanks{This work was partly funded by Huawei Sweden Technologies AB, Lund.}
	\thanks{
		The authors are with Lehrstuhl f\"ur Methoden der Signalverarbeitung, Technische Universit\"at M\"unchen, 80333 M\"unchen, Germany  (e-mail: benedikt.fesl@tum.de; nurettin.turan@tum.de; benedikt.boeck@tum.de; utschick@tum.de).
		
	}
	
}

%

\maketitle

\begin{abstract}
This work introduces a novel class of channel estimators tailored for coarse quantization systems. The proposed estimators are founded on conditionally Gaussian latent generative models, specifically \acp{gmm}, \acp{mfa}, and \acp{vae}. 
These models effectively learn the unknown channel distribution inherent in radio propagation scenarios, providing valuable prior information. 
Conditioning on the latent variable of these generative models yields a locally Gaussian channel distribution, thus enabling the application of the well-known Bussgang decomposition. 
By exploiting the resulting conditional Bussgang decomposition, we derive parameterized linear \ac{mmse} estimators for the considered generative latent variable models.
In this context, we explore leveraging model-based structural features to reduce memory and complexity overhead associated with the proposed estimators.
Furthermore, we devise necessary training adaptations, enabling direct learning of the generative models from quantized pilot observations without requiring ground-truth channel samples during the training phase. 
Through extensive simulations, we demonstrate the superiority of our introduced estimators over existing state-of-the-art methods for coarsely quantized systems, as evidenced by significant improvements in \ac{mse} and achievable rate metrics.
\end{abstract}


\IEEEpeerreviewmaketitle

\glsresetall
\begin{figure}[b]
	\onecolumn
	\centering
	\copyright \scriptsize{This work has been submitted to the IEEE for possible publication. Copyright may be transferred without notice, after which this version may no longer be accessible.}
	\vspace{-1.3cm}
	\twocolumn
\end{figure}
\section{Introduction}
\IEEEPARstart{M}{assive} \ac{mimo} and \ac{mmwave} systems enable the ever-increasing requirements of bandwidth and throughput in wireless communications. However, deploying a large number of high-precision \acp{adc} for each antenna's \ac{rf} chain with bandwidths sufficient for \ac{mmwave} systems is unaffordable in terms of cost and power consumption \cite{8758925,8337813}. One of the most direct and promising ways in order to solve the power consumption bottleneck and achieve high energy efficiency is to use low-resolution \acp{adc} at the \ac{bs}. In recent years, considerable research efforts have been devoted to analyzing the performance of low-resolution quantization systems 
\cite{7894211,ISIT2012_mezghani_nossek_ieee.pdf,Mezghani2016,5351659}.
Remarkably, although the low-resolution quantization causes nonlinear distortions at the receiver, the capacity is not severely reduced, especially at low \acp{snr} \cite{10.1145/1143549.1143827}.

In order to realize the mentioned favorable characteristics in practical low-resolution systems of the next generation of cellular systems (6G), accurate channel estimation is a crucial task. 
However, the severe nonlinearity of the \acp{adc} degrades the performance of conventional channel estimation algorithms~\cite{8758925,8337813}; thus, it is necessary to design novel channel estimators that provide good performance together with reasonable complexity and robustness in quantized systems.

Various channel estimation algorithms for coarsely quantized systems have been proposed in recent years.
In~\cite{6987288}, \ac{ls} estimation is considered, which is computationally simple but
results in rather poor estimation quality.
An iterative channel estimation technique based on the \ac{em} algorithm is proposed in \cite{5456454}, which exhibits limitations due to high complexity and convergence to local optima.
Iterative \ac{ml} methods are investigated in \cite{8683652,9641834}; however, they typically require a large number of pilot signals, resulting in an unaffordable signaling overhead~\cite{8337813}. 
In \cite{9358221,8586889,7355388}, joint channel estimation and decoding is investigated, where payload data is used to assist in channel estimation. Due to the iterative nature of the approaches, these methods are considered to have too high complexity for commercial massive \ac{mimo} systems \cite{8337813}. 
The works in \cite{qiht,8553303,Mo2018} take into account the sparsity of wireless channels and utilize \ac{cs} approaches such as iterative hard thresholding and \ac{gamp}; the main disadvantages thereby are the sensitivity concerning the (estimated) sparsity level and the high complexity of the iterative procedure.

\ac{mmse} channel estimation approaches in the case of a Gaussian channel with a diagonal covariance matrix are studied in \cite{8761531,5501995}; however, the \ac{mmse} estimator has no closed-form in the general case and is intractable to compute~\cite{fesl2023mean}. 
In fact, even the linear \ac{mmse} estimator has generally no closed-form solution. Only in the special case when the quantizer input is jointly Gaussian, the linear \ac{mmse} channel estimator can be efficiently computed based on Bussgang's theorem \cite{Bussgang} or the \ac{aqnm} \cite{4407763}, which is a special case of the Bussgang decomposition tailored to quantization~\cite{9307295}.\footnote{In this work, we use the terminology of the more general Bussgang decomposition, although the same results hold for the \ac{aqnm}}.
In \cite{Li2017,9500125}, the Bussgang estimator is derived for the one-bit as well as multi-bit quantization case.
On the one hand, it has the advantage that only a few pilot observations are necessary to achieve a good channel estimation quality. 
On the other hand, the estimator's applicability is seriously limited by the prerequisite of a Gaussian distributed channel with known second-order statistics.

Another recent branch of channel estimation techniques in quantized systems based on deep learning was investigated in 
\cite{9252921,9443563,Zhang2020,Dong2021}. 
Although providing good performance for low numbers of pilots, the approaches generally lack generalization ability with respect to different numbers of quantization bits, pilot signals, antennas, and \acp{snr}. 
Additionally, accumulating a large representative training dataset consisting of perfect \ac{csi} samples is necessary, which may be costly to acquire in practical systems, especially with coarse quantization.
In this paper, we address this major issue by deriving adapted training procedures for our models that enable learning directly from quantized training data without requiring any ground-truth \ac{csi} in the whole training phase.

In recent works, conditionally Gaussian latent generative models were utilized in order to learn the underlying unknown channel distribution of a radio propagation environment and leverage this prior information to design wireless communication functionalities \cite{turan2023versatile}, especially channel estimators for high-resolution systems \cite{9842343,9940363,10051921,fesl2023lowrank,baur2023leveraging,10051858}.
As mentioned before, the highly nonlinear distortion of low-resolution \acp{adc} makes it impractical to directly utilize the mentioned channel estimation techniques in coarsely quantized~systems. 
We extend this class of channel estimators based on conditionally Gaussian latent generative models by utilizing a novel proposed connection between these models and the possibility of using the linear \ac{mmse} estimator based on a newly proposed conditional version of the Bussgang decomposition.

Our contributions are concisely summarized as follows.
\begin{enumerate}
    \item We establish a theoretical foundation of extending the conventional Bussgang estimator to arbitrary channel \acp{pdf} by introducing conditional events such that the channel becomes conditionally Gaussian. Through this, we derive an approximation of the \ac{mse}-optimal \ac{cme} via a conditional Bussgang estimator that is the linear \ac{mmse} estimator. To find the conditional events of interest, we establish a novel connection to the topic of \ac{vi}, i.e., conditionally Gaussian latent models.

    \item 
    We introduce three variants of conditionally Gaussian latent models, i.e., \acp{gmm}, \acp{mfa}, and \acp{vae} that yield a tractable formulation of the proposed parameterized conditional Bussgang channel estimator that are linear \ac{mmse} estimators in quantized systems. Our work showcases the diverse strengths of the presented models concerning estimation performance, computational complexity, and memory overhead. 
    Notably, all three models share the advantage of being adaptable to different \acp{snr}, pilot sequences, and quantization bits without requiring re-training.

    \item 
    To facilitate practical feasibility, we propose training adaptations that allow us to learn the corresponding models solely from quantized pilot observations as training data, which eliminates the need for perfect \ac{csi} in the training stage. This feature enables us to use quantized pilot observations collected during the regular BS operation for training.
    In the case of the \ac{gmm}, we introduce a novel covariance recovery method that serves as an unbiased and consistent estimator of the unquantized input covariance matrix, using only quantized samples.
    Additionally, for the \ac{vae}, we adapt the \ac{elbo} loss function through model-based insights, specifically accounting for the nonlinear quantization process.
    
    \item 
    Extensive simulations based on realistic channel models demonstrate the strong performance of the proposed estimation framework, especially in setups with low numbers of pilot observations, surpassing state-of-the-art methods.
    
\end{enumerate}

\textit{Notation:} The standard Gaussian \ac{cdf} is denoted by $\op \Phi(x) = \int_{-\infty}^x \tfrac{1}{\sqrt{2\pi}}\exp(-\tfrac{x^2}{2})\op d x$ and the error function is given as $\erf(x) = \tfrac{2}{\sqrt{\pi}}\int_{0}^x \exp(-t^2)\op dt$. We denote the indicator function as $\chi(x \in \mathcal{A})$, which returns one if $x\in \mathcal{A}$ and zero otherwise. The $n$th entry of a vector is denoted by $[\B x]_n$. 
Conditional cross-covariance matrices are denoted as $\B C_{\B x\B y | \B c} = \op E[(\B x - \op E[\B x | \B c]) (\B y - \op E[\B y | \B c])\h | \B c]$, including the unconditional case $\B C_{\B x\B y}$, and auto-covariance matrices are abbreviated as $\B C_{\B x | \B c} = \B C_{\B x \B x | \B c}$.
The diagonals and off-diagonals of a matrix are denoted by $\diag(\B A)$ and $\nondiag(\B A) = \B A - \diag(\B A)$. We denote the Moore-Penrose inverse of a matrix $\B A$ as $\B A^\dagger$.

\section{Preliminaries}
\subsubsection{System Model}\label{subsec:system_model}
We consider the uplink transmission of $P$ pilot signals from a single-antenna \ac{mt} to an $N$-antenna \ac{bs} which operates \acp{adc} with $B$ quantization bits. The quantized receive signal is therefore written as
$\B R = Q_B(\B Y) = Q_B(\B h \B a\T + \B N)$,
where $\B R = [\B r_1,\dots, \B r_P]\in \mathbb{C}^{N \times P}$ contains the $P$ quantized receive signals as columns, $\B Y\in \C^{N\times P}$ describes the unquantized receive signal, $\B h\in \mathbb{C}^{N}$ denotes the wireless channel, $\B a\in\mathbb{C}^P$ is the pilot vector which fulfills the power constraint $\|\B a\|_2^2=P$, $\B N= [\B n_1,\dots, \B n_P]\in \mathbb{C}^{N\times P}$ is \ac{awgn} with $\B n_i\sim \NC(\B 0, \sigma^2\eye)$, and $Q_B$ denotes the $B$-bit quantization function, which is discussed below.
By column-wise vectorization, the system model can be written as
\begin{align}
    \B r = Q_B(\B y) = Q_B(\B A\B h + \B n) \in \C^{N P}
    \label{eq:sytem_model}
\end{align}
with $\B r  = \vect(\B R)$, $\B y  = \vect(\B Y)$, $\B n  = \vect(\B N)$, and $\B A = \B a \otimes \eye$. 
Note that an extension to a multi-user setup can, in principle, be straightforwardly achieved by stacking the channels of all users, cf., e.g., \cite{Li2017}; however, the analysis of multi-user systems is out of the scope of this work.
By normalizing the channels as $\op E[\|\B h\|_2^2] = N$, the \ac{snr} of the quantizer input is defined as $\text{SNR} = 1/\sigma^2$.

Typically, several pilot observations are required to achieve reasonable channel estimation performance in coarsely quantized systems.
For the case of one-bit quantization, it is shown in \cite{fesl2023mean} that a pilot sequence with equidistant phase shifts in the range $[0, \frac{\pi}{2})$ is \ac{mse}-optimal with respect to the \ac{cme} for jointly Gaussian inputs in the asymptotic high \ac{snr} regime.
In the general case, the optimal pilot sequence depends on many system parameters and is thus intractable to derive.
Therefore, in this work, we consider pilots that have an equidistant spacing in both the amplitude and the angle,~i.e.,
\begin{align}\label{eq:pilot_vec}
    [\tilde{\B a}]_i = \beta_i \exp\left(\op j \frac{\pi}{2P}(i-1)\right),~i\in\{1,\dots,P\},
\end{align}
where $\beta_i = \frac{1}{2} + \frac{i-1}{2(P-1)}$ is the amplitude spacing. In order to fulfill the power constraint $\|\B a\|_2^2 = P$, the pilot vector is normalized as $\B a = \frac{\sqrt{P}}{\|\tilde{\B a}\|_2}\tilde{\B a}$. 
The choice in \eqref{eq:pilot_vec} has shown to be robust with respect to all considered scenarios in our simulations.
However, the presented class of estimators is independent of the specific pilot sequence and can be deployed with any other desired pilot sequence of choice.

\subsubsection{Quantizer Design}\label{subsec:quantizer_design}

In this work, we consider scalar quantizers, i.e., the quantization is performed elementwise on the input, where the real and imaginary parts are quantized independently. The quantizer can be described by means of the $2^B$ quantization labels $\ell_i$, $i\in\left\{1,\dots, 2^B\right\}$, and the quantization thresholds $\tau_i$, $i\in \left\{0,\dots, 2^B\right\}$, where $\tau_{0} = -\infty$ and $\tau_{2^B} = \infty$ by definition.
The quantization function of the real/imaginary part of the signal can be denoted as 
\begin{align}\label{eq:quantizer}
    Q_B(x) = \sum_{i=1}^{2^B} \ell_i\hspace{1pt}\chi(\tau_{i-1} \leq x < \tau_i).
\end{align}
For the case of one-bit quantization $B=1$, the quantization function of the complex-valued signal $\B y$ can be expressed as
\begin{align}
    Q_{1}(\B y) = \frac{1}{\sqrt{2}}\left(\sign(\Re(\B y)) + \op j \sign(\Im(\B y))\right).
\end{align}

Practicable \acp{adc} usually have uniformly spaced quantization thresholds with a constant step size $\Delta$, which depends on the input distribution, and the quantization labels are placed in the middle of two quantization thresholds. For the case of zero-mean Gaussian input with variance one, the optimal values for $\Delta$ are computed numerically in \cite{1057548}. 

Under the assumption that the elementwise quantizer input is zero-mean Gaussian distributed with variance $1+\sigma^2$ following the considered SNR definition, we choose the SNR-dependent step size as suggested in \cite{9448919} as
\begin{align}\label{eq:step_uniform}
    \Delta = \sqrt{\tfrac{1}{2}(1 + \sigma^2)} \Delta_*
\end{align}
where $\Delta_*$ is the step size for the standard Gaussian input~\cite{1057548}.
Although the quantizer input is generally not Gaussian distributed, this choice gives a reasonable performance with regard to practical feasibility.
The necessary scaling in \eqref{eq:step_uniform} is resolved by automatic gain control in practice.
We note that there exist more sophisticated quantizer designs, e.g., non-uniform scalar quantization \cite{1057548,720541}; 
however, the uniform quantizer is considered the most practicable choice in wireless communications \cite{Mezghani2016}. Furthermore, the channel estimation techniques proposed in this work 
can be straightforwardly extended to non-uniform quantization.

\subsubsection{Channel Models}\label{sec:channel_model}

We work with the \ac{3gpp} spatial channel model~\cite{3gpp,NeWiUt18} where channels are modeled conditionally Gaussian: $\B h| \B \delta \sim \NC(\B 0, \B C_{\B\delta})$.
The random vector $\B \delta$ collects the angles of arrival/departure and path gains of the main propagation clusters between a \ac{mt} and the \ac{bs}.
The main angles are drawn independently and uniformly from the interval $[0, 2\pi]$; the path gains are also drawn uniformly and are subsequently normalized such that they sum up to one.
The \ac{bs} employs a \ac{ula} such that the spatial channel covariance matrix is given by
\begin{equation}\label{eq:3gpp_cov}
    \B C_{\B h| \B \delta} = \int_{-\pi}^\pi \omega(\gamma; \B \delta) \B t(\gamma) \B t(\gamma)\h \op d \gamma.
\end{equation}
Here,
    $\B t(\gamma) = [1, \op e^{\op j\pi\sin(\gamma)}, \dots, \op e^{\op j\pi(N-1)\sin(\gamma)}]\T$
is the array steering vector for an angle of arrival $\gamma$, and $\omega$ is a power density consisting of a sum of weighted Laplace densities whose standard deviations describe the angle spread of the propagation clusters~\cite{3gpp}.
For every channel sample, we generate random angles and path gains, combined in $\B \delta$, and then draw the sample as $ \B h \sim \NC(\B 0, \B C_{\B h | \B \delta}) $, which results in an overall non-Gaussian channel distribution \cite{9842343}. Note that the conditional Gaussianity of the channel model is not connected to the conditional Gaussianity of the proposed latent models since the inference of \eqref{eq:3gpp_cov} from a single snapshot is intractable.


To ensure a broader evaluation of different channel models, version 2.4 of the QuaDRiGa channel simulator \cite{QuaDRiGa1,dbt_mods_00032895} is used to generate channel samples. 
We simulate an urban macrocell scenario at a center frequency of 6 GHz.
The \ac{bs}'s height is 25 meters, and it covers a $120^\circ$ sector.
The distances between the \acp{mt} and the \ac{bs} are in the range of 35--500 meters. We either consider a pure \ac{los} scenario or a mixed \ac{los}/\ac{nlos} scenario, where in 80\% of the cases, the \acp{mt} are located indoors at different floor levels, whereas the \acp{mt}' height is 1.5 meters in the case of outdoor locations.
The \ac{bs} is equipped with a \ac{ula} with $N$ ``3GPP-3D'' antennas, and the \acp{mt} employ an omnidirectional antenna. The generated channels are post-processed to remove the effective path gain \cite[Sec. 2.7]{dbt_mods_00032895}.

\subsubsection{Training Datasets}\label{subsec:datasets}
In this work, we consider the channel distribution to be unknown and arbitrarily complex by means of the sophisticated channel models, cf. \Cref{sec:channel_model}. However, we assume the availability of a representative dataset which is comprised of samples stemming from the respective channel model. In practice, this means that data samples from the respective \ac{bs} cell are available. In this work, we discuss two different setups. First, we assume the availability of a training dataset consisting of $T$ ground-truth channel samples $\mathcal{H} = \{\B h_t\}_{t=1}^T$. This can be achieved in practice via measurement campaigns or digital twins, e.g., ray tracing. Afterward, we consider a training dataset consisting solely of noisy and quantized pilot observations $\mathcal{R} = \{\B r_t\}_{t=1}^T$. Therefore, pilot observations from the regular \ac{bs} operation can be utilized, and no ground-truth channel samples are needed.

\section{Bussgang Estimator}\label{sec:Bussgang}
In this section, we briefly revise the linear \ac{mmse} estimator based on the Bussgang decomposition, which is a direct consequence of Bussgang's theorem \cite{Bussgang}. 
As stated above, an equivalent derivation can be done via the \ac{aqnm}, cf.~\cite{9307295}.
Although the Bussgang decomposition generally exists, the Bussgang linear \ac{mmse} estimator is analytically tractable only if the channel and noise follow a zero-mean Gaussian distribution \cite{Li2017}.
However, even if this is generally not true, an approximation to the linear \ac{mmse} estimator, assuming the channel is zero-mean Gaussian, is a reasonable baseline for channel estimation in quantized~systems.

In particular, under the assumption of a jointly zero-mean Gaussian quantizer input, the Bussgang decomposition implies that the system in \eqref{eq:sytem_model} can be written as a linear combination of the desired signal part and an uncorrelated distortion $\B q$ as
\begin{equation}
	\B r = Q_B(\B y) = \B B \B y + \B \eta = \B B\B A\B h + \B q,
	\label{eq:Bussgang_decomp}
\end{equation}
where $\B B$ is the Bussgang gain that can be obtained from the linear \ac{mmse} estimation of $\B r$ from $\B y$ as $\B B = \B C_{\B r \B y}\B C_{\B y}\inv$, 
cf.~\cite[Sec. 9.2]{Papoulis1965ProbabilityRV}, and where the distortion term $\B q = \B B \B n + \B \eta$ contains both the \ac{awgn} $\B n$ and the quantization noise $\B \eta$.
The Bussgang gain matrix for a uniform quantizer with jointly Gaussian input is derived in~\cite{7967843} and is computed as
\begin{equation}\label{eq:Buss_multi}
    \begin{aligned}
        \B B = \frac{\Delta}{\sqrt{\pi}}\B D_{\B y}^{-\frac{1}{2}} 
        \sum_{i=1}^{2^B-1} \exp\left(-\Delta^2\left(i - 2^{B-1}\right)^2\B D_{\B y}\inv \right)
    \end{aligned}
\end{equation}
where $\B D_{\B y} = \diag(\B C_{\B y})$.
In the case of one-bit quantization $B=1$, by choosing $\Delta=\sqrt{2}$, we get the well-known solution
\begin{align}\label{eq:Buss_one}
    \B B = \sqrt{\frac{2}{\pi}}\diag(\B C_{\B y})^{-\frac{1}{2}}.
\end{align}

As the statistically equivalent model \eqref{eq:Bussgang_decomp} is linear, one can formulate the linear \ac{mmse} estimator 
\begin{equation}
	\hat{\B h}_{\text{Buss}} = \B C_{\B h\B r} \B C_{\B r}\inv\B r.
	\label{eq:h_buss}
\end{equation}
The cross-correlation matrix between the channel and the received signal is calculated as
$\B C_{\B h\B r} = \op{E}[\B h(\B B \B A \B h + \B q)\h] = \B C_{\B h}\B A\h\B B\h$
which follows from the fact that the noise term $\B q$ is uncorrelated with the channel $\B h$, see \cite[Appendix A]{Li2017}. Note that this property only holds in the case of a jointly Gaussian quantizer input.
For the one-bit quantization case, the auto-correlation matrix is equal to the covariance matrix $\B C_{\B r}$ due to the elimination of the amplitude information and can be calculated in closed-form via the so-called arcsine law \cite{272490} as 
\begin{equation}\label{eq:arcsin_law}
	\begin{aligned}
		\B C_{\B r} = \frac{2}{\pi}\left( \arcsin\left(\B D_{\B y}^{-\frac{1}{2}}\Re(\B C_{\B y})\B D_{\B y}^{-\frac{1}{2}}\right) \right.
		\\
		\left.
		+ \op j \arcsin\left(\B D_{\B y}^{-\frac{1}{2}}\Im(\B C_{\B y})\B D_{\B y}^{-\frac{1}{2}}\right)\right).
	\end{aligned}
\end{equation}

Unfortunately, for the multi-bit quantization case, no closed-form expression for $\B C_{\B r}$ exists. 
Besides that, the computation of the variances after the quantization are no longer non-trivial.
As shown in \cite[eq. (2.14)]{Mezghani2016} (adapted for the complex-valued case), for Gaussian input and the uniform quantizer, the variances are computed as
\begin{equation}\label{eq:quant_variance}
    [\B C_{\B r}]_{i,i} =  \sum_{i=1}^{2^B} 2\ell_i^2 \left(\op \Phi\left(\sqrt{2}\tau_{i}c_i \right) - \op \Phi\left(\sqrt{2}\tau_{i-1}c_i\right) \right),
\end{equation}
where $c_i = [\B C_{\B y}]_{i,i}^{-\frac{1}{2}}$.
Although there exist various practicably feasible approximations for the evaluation of the involved Gaussian \ac{cdf}, cf. \cite{Cody1969,1094433}, the evaluation of \eqref{eq:quant_variance} may still be problematic in time-critical systems; thus, a reasonable approximation is used in this work. By assuming that the signal's variance does not change for different antennas, the Bussgang gain becomes a scaled identity of the form $\B B = \rho \eye$, cf. \eqref{eq:Buss_multi}. By further neglecting cross-correlations of the quantization distortion, the quantized covariance matrix $\B C_{\B r}$ is well approximated, especially in the low \ac{snr} regime, by, cf. \cite{7894211},
\begin{align}\label{eq:cr_multibit}
    \B C_{\B r} \approx \rho^2 \B C_{\B y} + (1 - \rho^2) \diag(\B C_{\B y}).
\end{align}
Importantly, the resulting covariance matrix $\B C_{\B r}$ remains \ac{psd} if $0\leq \rho^2 \leq 1$.
We note that the expression for $\B C_{\B r}$ with respect to $\B C_{\B y}$ generally depends on the quantizer choice and the input distribution, and useful approximations can be found differently, cf., e.g., \cite{ISIT2012_mezghani_nossek_ieee.pdf}. However, the design of the channel estimation algorithms in this work is not founded upon the choice in \eqref{eq:cr_multibit}, and different approximations can be utilized.


\section{Parameterized Bussgang Channel Estimators}\label{sec:main_part}

The prerequisite of the Bussgang estimator in \Cref{sec:Bussgang} that the quantizer input is jointly Gaussian imposes a severe limitation. This becomes especially evident when considering the channel distribution of a whole \ac{bs} cell, which is strongly shaped by the propagation environment, generally being considerably underrepresented by a simple Gaussian distribution. 
Although the Bussgang decomposition in principle also exists for the non-Gaussian case, the corresponding Bussgang gain matrix is non-diagonal and not analytically tractable, which, in turn, yields no analytic solution for the linear \ac{mmse} estimator \cite{9307295}. This motivates us to utilize the powerful concept of conditional Gaussianity, which is already used for channel estimation in high-resolution systems \cite{9842343,9940363,10051921,fesl2023lowrank,baur2023leveraging,10051858}.
In the following Lemma \ref{lemma:cond_Buss}, we establish the theoretical foundation for the estimation framework in quantized systems through a \textit{conditional} Bussgang decomposition.

\begin{lemma}
\label{lemma:cond_Buss}
    Consider the system model \eqref{eq:sytem_model} with a uniform quantizer $Q_B(\cdot)$ and let $\B h \sim p(\B h)$ with an arbitrary \ac{pdf} $p(\B h)$.
    Let  $\B c$ be a conditional event independent of $\B n$ such that
    \begin{equation}
        \B h |\B c \sim \mathcal{N}_\C(\B h; \B 0, \B C_{\B h|\B c})
        \label{eq:conditional_Gaussian_general}
    \end{equation}
    and $\B C_{\B y|\B c} = \B A \B C_{\B h|\B c} \B A\h + \sigma^2\eye$.
    Then, there exists a unique conditional Bussgang gain $\B B_{\B c}$ such that \eqref{eq:sytem_model} can be decomposed as the statistically equivalent model
    \begin{equation}
        \B r  = \B B_{\B c}\B y + \B \eta = \B B_{\B c} \B A \B h + \B B_{\B c} \B n + \B \eta
        \label{eq:conditional_Bussgang_decomp_general}
    \end{equation} 
    where $\B \eta$ and $\B h$ are conditionally uncorrelated given $\B c$ with the following properties:
    \begin{itemize}
        \item[a)] $\B B_{\B c}$ is computed in closed-form via \eqref{eq:Buss_multi} (or \eqref{eq:Buss_one} for $B=1$), 
        
        \item[b)] $\B C_{\B r | \B c}$ is computed in closed-form via \eqref{eq:arcsin_law} for $B=1$, 
            
        \item[c)] the diagonal entries of $\B C_{\B r | \B c}$ are computed via \eqref{eq:quant_variance} and $\B C_{\B r | \B c}$ is well approximated by \eqref{eq:cr_multibit} for $B>1$,

    \end{itemize}
        where $\B C_{\B y}$ is substituted with $\B C_{\B y|\B c}$ in \eqref{eq:Buss_multi}, \eqref{eq:Buss_one}, and \eqref{eq:arcsin_law}--\eqref{eq:cr_multibit}.
    Further, the conditional linear \ac{mmse} estimator of $\B h$ given $\B r$ and $\B c$ is computed as 
    \begin{align}
        \hat{\B h}_{\B c}(\B r) = \B C_{\B h|\B c}\B A\h \B B_{\B c}\h \B C_{\B r|\B c}\inv \B r.
        \label{eq:cond_lin_MMSE}
    \end{align}
\end{lemma}

\textit{Proof:} See Appendix \ref{app:proof_lemma1}.

The main result of Lemma \ref{lemma:cond_Buss} shows that the Bussgang estimator from \Cref{sec:Bussgang} can be extended to arbitrary channel \acp{pdf} when finding conditional events such that the channel becomes (conditionally) zero-mean Gaussian. Although it is a promising result, finding such conditional events is generally highly non-trivial, especially because the conditional event must not be a function of the observation, as this would introduce a dependency with respect to the noise realization.

However, when modeling the conditional event $\B c$ as a latent random variable with a prior distribution $p(\B c)$ of choice and enforcing \eqref{eq:conditional_Gaussian_general} via a parameterized distribution, aiming to maximize the likelihood $p(\B h)$ with a given dataset, we end up in the classical framework of \ac{vi} \cite[Ch. 10]{bookBi06} in the special case of conditionally Gaussian latent variable models.
The Bayesian modeling of the conditional event further allows us to approximate the \ac{mse}-optimal \ac{cme}, which is generally intractable \cite{fesl2023mean}, as
\begin{align}
    \op E[\B h |\B r] &= \op E[\op E [\B h | \B r, \B c]| \B r] 
    = \int \op E [\B h | \B r, \B c] p(\B c | \B r) \dx \B c 
    \label{eq:cme1}
    \\
    &\approx \int \hat{\B h}_{\B c}(\B r) p(\B c | \B r) \dx \B c
    \label{eq:cme2}
\end{align}
where in \eqref{eq:cme1} we use the law of total expectation and in \eqref{eq:cme2} we approximate the conditional mean $\op E [\B h | \B r, \B c]$ by the linear \ac{mmse} estimate \eqref{eq:cond_lin_MMSE} derived in Lemma \ref{lemma:cond_Buss}. Note that in \eqref{eq:cme1}, we have implicitly assumed that the latent variable is continuous, albeit discrete latent variables can equivalently be used, as shown later.
Although \eqref{eq:cme2} remains an approximation, the linear \ac{mmse} estimator is widely adopted for channel estimation, especially because it allows for a low-cost implementation due to the desirable linearity of the filter. The marginalization over the latent variable $\B c$ in \eqref{eq:cme2} remains to be solved in a tractable manner by the \ac{vi} formalism of choice, for which we discuss several variants in the remainder of this section, including the \ac{gmm}, the \ac{mfa}, and the \ac{vae}. 
Throughout this section, it is assumed that a training dataset $\mathcal{H}$ of ground-truth channel samples is available, cf. \Cref{subsec:datasets}.

\subsection{GMM-based Bussgang Estimator}\label{subsec:gmm}
We start by deriving the \ac{gmm}-based estimator, which parameterizes a \textit{componentwise} Bussgang estimator. Generally, a \ac{gmm} is a \ac{pdf} of the form 
\begin{align}
    p^{(K)}(\B h) = \sum_{k=1}^K \pi_k \NC(\B h; \B \mu_{\B h| k}, \B C_{\B h|k})
\end{align}
where $K$ is the number of mixture components and $\{\pi_k,\B \mu_{\B h|k}, \B C_{\B h| k}\}_{k=1}^K$ is the set of parameters of the \ac{gmm}, namely the mixing coefficients, the means, and the covariances of the Gaussian components.
The parameters of the \ac{gmm} are fitted via the \ac{em} algorithm for a given training dataset $\mathcal{H}$ of channel samples, cf. \cite[Ch. 9]{bookBi06}.
An essential property of \acp{gmm} is that for a given data sample, the \textit{responsibility} of each component can be computed as, cf. \cite[Ch. 9]{bookBi06}, $p(k | \B h) \propto \pi_k \NC(\B h; \B \mu_{\B h| k}, \B C_{\B h|k})$.
The \ac{gmm} can be described via a \textit{discrete} latent variable with a categorical distribution, which conditions on one of the $K$ components \cite[Ch. 9]{bookBi06} and, thus, yields a conditionally Gaussian latent variable model.

To ensure the validity of \eqref{eq:conditional_Gaussian_general} in Lemma \ref{lemma:cond_Buss}, we enforce the component means to be zero, i.e., $\B \mu_{\B h|k} = \B 0$ for all $ k\in \{1,\dots,K\}$. To reflect this constraint in the fitting process, the component means are set to zero in every M-step of the \ac{em} algorithm. 
Naturally, the zero-mean constraint diminishes the capability of the \ac{gmm} to a certain extent concerning its ability to approximate the true underlying distribution. 
Nevertheless, since a feasible wireless channel distribution is considered to be zero-mean with a decreasing probability density towards higher amplitudes, cf., e.g., \cite{3gpp}, the loss of accuracy of the model can be considered to be small.
Moreover, restricting the component means prevents overfitting and allows to model high-dimensional data \cite{pan07a}.

After the \ac{gmm} is trained, it is used for channel estimation similar to \cite{9842343} but with multiple adaptions in order to take the quantization effect into account, as outlined in Lemma~\ref{lemma:cond_Buss}. 
We first note that if the channel distribution is modeled as a zero-mean \ac{gmm}, also the distribution of the unquantized receive signal $\B y$ follows a zero-mean \ac{gmm} with covariances $\B C_{\B y|k} = \B A\B C_k\B A\h + \sigma^2\eye$ for all $ k\in\{1,\dots,K\}$, cf. \eqref{eq:sytem_model}. 
For each \ac{gmm} component, we can apply the conditional Bussgang decomposition to find a statistically equivalent model, cf. \eqref{eq:conditional_Bussgang_decomp_general} in Lemma \ref{lemma:cond_Buss}:
\begin{align}
    \B r = \B B_k \B A\B h + \B q_k,
\end{align}
where $\B B_k$ is the conditional Bussgang gain of component $k$, and $\B q_k = \B B_k \B n + \B \eta$. 
According to Lemma \ref{lemma:cond_Buss}, the computation of the conditional Bussgang gain $\B B_k$ is done via the closed-form solutions in \eqref{eq:Buss_multi} or \eqref{eq:Buss_one}, respectively.

We note that the integral in \eqref{eq:cme2} to approximate the \ac{cme} simplifies to a sum because of the discrete latent variable of the \ac{gmm}. 
Opposite to the high-resolution case, the evaluation of the discrete distribution $p(\B r|k)$ is intractable in general since the cardinality of its discrete support increases exponentially in the number of dimensions \cite{fesl2023mean}.
Thus, in order to evaluate the responsibility for a given pilot signal, we assume that the quantized receive signal follows a zero-mean \ac{gmm} distribution with the same second-order moments; this assumption effectively resembles approximate inference~\cite[Ch. 10]{bookBi06}.
The covariance matrix of component $k$, named $\B C_{\B r|k}$, is thereby computed via \eqref{eq:arcsin_law} or \eqref{eq:cr_multibit} for the one- or multi-bit quantization case, respectively, by plugging in the component's unquantized covariance $\B C_{\B y|k}$, cf. Lemma \ref{lemma:cond_Buss}. 
Thereby, in the case of a covariance matrix $\B C_{\B h|k}$ with a non-constant diagonal, the scaling parameter $\rho_k$ in \eqref{eq:cr_multibit} for the $k$th component is approximated via $\rho_k = \min(\frac{1}{N} \sum_{i=1}^N [\B B_k]_{i,i}, 1)$, which ensures that the resulting matrix is \ac{psd}.
This yields the following responsibility evaluation of the quantized receive signal:
\begin{align}\label{eq:resp_r}
    p(k | \B r) \approx \frac{\pi_k \NC(\B r; \B 0, \B C_{\B r|k})}{\sum_{i=1}^K \pi_i \NC(\B r; \B 0, \B C_{\B r|i})}.\end{align}

The final channel estimate is computed via the convex combination of the componentwise Bussgang estimators, 
which follows from \eqref{eq:cme2} together with Lemma \ref{lemma:cond_Buss},
parameterized by the \ac{gmm} covariances, which yields
\begin{align}\label{eq:BGMM}
    \hat{\B h}_{\text{BGMM}}^{(K)}(\B r) = \sum_{k=1}^K p(k| \B r) \B C_{\B h|k} \B A\h \B B_k\h \B C_{\B r|k}\inv \B r.
\end{align}

We note that the \ac{gmm} with fully parameterized covariance matrices is independent of any array geometries at the \ac{bs}.
However, as shown in \cite{10051921}, it is possible to enforce different structural constraints for the \ac{gmm}'s covariances, such as a circulant (``GMM circ'') or Toeplitz (``GMM toep'') structure.
These structural constraints reflect the typical array geometries of a \ac{bs}, e.g., a \ac{ula} or \ac{upa}, and result in a reduced number of parameters and a lower online complexity of the estimator due to the usage of 1D or 2D \acp{fft} \cite{10051921}, respectively. 
The covariance matrix of the $k$th \ac{gmm} component is thereby constrained to be of the form $\B C_{\B h|k} = \B Q\h\diag(\B c_{\B h|k})\B Q$
where $\B Q$ is an (oversampled) 1D or 2D \ac{dft} matrix in the case of a \ac{ula} or \ac{upa}, respectively, and $[\B c_{\B h|k}]_i\in \mathbb{R}_+$.
The structural constraints are, without limitation, also applicable in coarsely quantized systems since the quantization is performed elementwise on the input and thus does not alter the imposed array structure.
In this work, we consider solely the case of a \ac{ula} as mentioned in \Cref{sec:channel_model}.
The necessary memory overhead and computational complexity of the resulting estimators are discussed in more detail in \Cref{sec:memory_complexity}.

\subsection{MFA-based Bussgang Estimator}\label{subsec:mfa}
A related concept to the \ac{gmm} is the \ac{mfa} model, which, in addition to a discrete latent variable $k$ which describes the mixture component, also contains a continuous latent variable $\B z\in \mathbb{C}^L$ of lower dimension, i.e., $L < N$ holds \cite[Ch. 12]{Murphy2012}, \cite{fesl2023lowrank}.
This effectively models the data on a piecewise linear subspace. 
After integrating out the continuous latent variable $\B z\sim \N(\B 0, \eye)$, the \ac{pdf} of the \ac{mfa} model is a special form of a \ac{gmm} with low-rank plus diagonal-constrained covariances of the form 
\begin{align}
    p^{(K,L)}(\B h) = \sum_{k=1}^K \pi_k \mathcal{N}_{\mathbb{C}}(\B h; \B \mu_{\B h|k}, \B W_{\B h|k}\B W_{\B h|k}\h + \B \Psi_{\B h|k}),
    \label{eq:mfa}
\end{align}
where $\B W_{\B h|k}\in \mathbb{C}^{N\times L}$ is the \textit{factor loading} matrix and $\B \Psi_{\B h|k} \in \mathbb{C}^{N\times N}$ is a diagonal matrix. 
In order to fit the parameters $\{\pi_k, \B \mu_{\B h|k}, \B W_{\B h|k}, \B \Psi_{\B h|k}\}_{k=1}^K$ of the \ac{mfa} model for a given dataset $\mathcal{H}$ of channel realizations, an \ac{em} algorithm can be used \cite[Ch. 12]{Murphy2012}. 
After training, by defining $\B C_{\B h|k} = \B W_{\B h|k}\B W_{\B h|k}\h + \B \Psi_{\B h|k}$, the model can be effectively treated as a \ac{gmm}.
A zero-mean \ac{mfa} model with $\B \mu_{\B h|k} = \B 0$  for all $k\in \{1,\dots,K\}$ can be similarly enforced as in the \ac{gmm} case.
Similar to \cite{fesl2023lowrank}, we set $\B \Psi_{\B h | k} = \psi_{\B h | k}\eye$ for all $ k\in\{1,\dots,K\}$.

The main advantage of the \ac{mfa} model in the context of channel estimation, in contrast to the \ac{gmm}, lies in the reduced number of parameters due to the low-dimensional latent space, which mitigates overfitting effects during training. Thus, it is a more robust model for lower numbers of training data, as demonstrated in \cite{fesl2023lowrank}. 
Since the \ac{mfa} also parameterizes a conditionally Gaussian distribution, we can apply Lemma \ref{lemma:cond_Buss}, resulting in a componentwise Bussgang estimator similar to the \ac{gmm} case, see \Cref{subsec:gmm}, which yields the \ac{mfa}-parameterized Bussgang channel estimator $\hat{\B h}_{\text{BMFA}}$ by substituting the respective covariances in \eqref{eq:BGMM}.
We note that the \ac{mfa} model enforces the covariance structure via training and is thus independent of the \ac{bs}'s array geometry.

\subsection{VAE-based Bussgang Estimator}\label{subsec:vae}

\begin{figure}[t]
    \centering
    \resizebox{1\columnwidth}{!}{
        \includegraphics{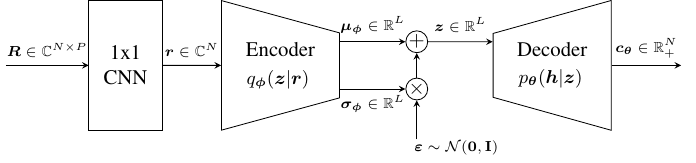}
    }
    \caption{Proposed adapted \ac{vae} architecture with the encoder, latent space, and decoder together with the parameterized distributions and the reparameterization trick.}
    \label{fig:vae_architecture}
\end{figure}

The \ac{vae} was introduced in \cite{Kingma2014} and has attracted a lot of interest in the area of generative modeling due to its strong performance, which builds on the basis of \acp{nn} that are used for the encoder and decoder of the \ac{vae}, cf. Fig. \ref{fig:vae_architecture}. In \cite{baur2023leveraging, 10051858}, the \ac{vae} was successfully utilized for channel estimation in high-resolution systems.
In contrast to the \ac{gmm} and \ac{mfa}, the \ac{vae} comprises a continuous and \textit{nonlinear} latent space, encoded by the low-dimensional latent vector $\B z\in \mathbb{R}^L$.
The most common design choice is a Gaussian model for the latent vector, i.e., $\B z \sim \N(\B 0, \eye)$.
Since the variational inference task is no longer tractable by the classical \ac{em} algorithm, \acp{nn} in combination with the reparameterization trick are used to train the \ac{vae} \cite{Rezende2014,Kingma2014}.
In this work, we choose the parameterized encoder and decoder distributions as
\begin{align}
    q_{\phi}(\B z | \B r) &= \N(\B z; \B \mu_{\phi}(\B r), \diag(\B \sigma_{\phi}^2(\B r))),
    \label{eq:vae_enc}
    \\
    p_{\theta}(\B h | \B z) &= \NC(\B h; \B 0, \dft\h \diag(\B c_{\theta}(\B z)) \dft),~\B c_{\theta}\in\mathbb{R}_+^N,
    \label{eq:vae_dec}
\end{align}
respectively. Note that \eqref{eq:vae_dec} is chosen to fulfill \eqref{eq:conditional_Gaussian_general} in Lemma~\ref{lemma:cond_Buss}.
The matrix $\dft$ is a \ac{dft} matrix such that the parameterized channel covariance matrix of the \ac{vae} is a circulant matrix, cf. \cite{10051858}. 
This choice results in a reduced number of parameters and is justified by the imposed structure of the \ac{ula} at the \ac{bs}, similar to the \ac{gmm} case, cf. \Cref{subsec:gmm}. The extension to the \ac{upa} case is once again straightforward by replacing the 1D \ac{dft} matrix with its 2D counterpart.

For every training data point $\B h_t \in \mathcal{H}$, the \ac{vae} computes the \ac{elbo} on the log-likelihood \cite{Kingma2014}
\begin{align*}
    \log p_{\theta}(\B h_t) \geq \op E_{\phi}[\log p_{\theta}(\B h_t | \B z)] - \DKL(q_{\phi}(\B z | \B h_t)\hspace{1pt}||\hspace{1pt} p(\B z)).
\end{align*}
By plugging in \eqref{eq:vae_enc} and \eqref{eq:vae_dec} and ignoring the constant terms (they do not influence the optimization), the \ac{elbo} is utilized as the loss function $L_{\phi,\theta}$ of the \ac{vae}, which reads as, cf. \cite{10051858},
\begin{equation}\label{eq:elbo}
    \begin{aligned}
    L_{\phi,\theta}(\B h_t) = \sum_{n=1}^N -\log [\B c_{\theta}]_n - \B h\h\dft\h \diag(\B c_{\theta})\inv \dft\B h 
    \\
    + \sum_{\ell=1}^L \log [\B \sigma_{\phi}]_\ell - \frac{1}{2} [\B \mu_\phi]_\ell^2 - \frac{1}{2} [\B \sigma_\phi]_\ell^2.
    \end{aligned}
\end{equation}

Following Lemma \ref{lemma:cond_Buss}, the \ac{vae} can be utilized to parameterize a channel estimator by means of the conditional Gaussianity at the output of the decoder of the \ac{vae} in combination with linear \ac{mmse} filters.
After the training, we utilize the \ac{vae} to parameterize a channel covariance matrix for each quantized receive pilot $\B r$ by forwarding the pilot through the encoder and then using the latent mean $\B \mu_{\phi}(\B r)$ as input to the decoder which yields the channel covariance matrix $\B C_{\theta} = \dft\h \diag(\B c_{\theta}) \dft$. This procedure approximates \eqref{eq:cme2} with a low-complexity implementation, which has been shown to work well in the high-resolution case \cite{baur2023leveraging, 10051858}.
The \ac{vae}-parameterized Bussgang estimator then reads as 
\begin{equation}
    \hat{\B h}_{\text{BVAE}}(\B r) = \dft\h \diag(\B c_{\theta}) \dft \B A\h \B B_{\B z}\h \B C_{\B r|\B z}\inv \B r,
\end{equation}
where $\B B_{\B z}$ is computed by plugging $\B C_{\B y|\B z} = \B A\B C_{\theta}\B A\h + \sigma^2 \eye$ into \eqref{eq:Buss_multi} or \eqref{eq:Buss_one}. Similarly, the covariance $\B C_{\B r|\B z}$ is computed by plugging $\B C_{\B y|\B z}$ into \eqref{eq:arcsin_law} or \eqref{eq:cr_multibit}, cf. Lemma \ref{lemma:cond_Buss}.

For the encoder and decoder, we use a four-layer feed-forward \ac{nn} with \ac{relu} activation functions, respectively, for which we stack the real and imaginary parts of the pilot observation at the input to the encoder. In the case of multiple pilot observations, we add a \ac{cnn} with $P/2$ layers and \ac{relu} activation functions before the encoder, which perform $1\times 1$ convolutions in order to always have a $2N$-dimensional input at the encoder. This modification drastically reduces the number of parameters and simplifies a forward pass through the \ac{vae}. The complete \ac{vae} architecture is detailed in Fig. \ref{fig:vae_architecture}.

\section{Enabling Learning from Quantized Pilot Observations as Training Data}\label{sec:learning_quant}
As discussed above, the availability of a large training dataset $\mathcal{H}$ consisting of representative ground-truth channel samples for a whole \ac{bs} cell (radio propagation scenario) is questionable in practical communication systems. Although different approaches exist, such as ray tracing, to generate a training dataset that mimics the underlying channel distribution, it is unclear whether they can sufficiently capture the characteristics of a real communication scenario. A different idea is to train the respective models directly on pilot observations and mitigate imperfections, e.g., additive noise or sparsely allocated pilots, through model-based training adaptations. This has already been shown to work well for high-resolution scenarios \cite{10051858,10078293}.
However, learning a generative model from a training dataset $\mathcal{R}$ consisting of quantized data poses a significant challenge due to the pronounced nonlinear distortion resulting from low-resolution quantization.
In the remainder of this section, we propose two novel training adaptations to the \ac{gmm} and the \ac{vae} to learn the underlying channel distributions, although only quantized training data and no ground-truth \ac{csi} is available.

\subsection{Covariance Recovery for GMM Approximation}\label{subsec:gmm_quant}
Recovering the unquantized covariance matrix of the input to a one-bit quantizer solely from quantized data has gained a lot of interest very recently \cite{Dirksen21,xiao2023onebit,9930674}. 
Since the amplitude information is lost in the case of a zero-threshold one-bit quantization, only the normalized correlation matrix can be obtained via the inverse arcsine law \cite{272490}. To resolve this issue, non-zero or time-varying quantizer thresholds were considered in order to be able to estimate the variances of the input signal and, thus, the whole covariance matrix \cite{Dirksen21,xiao2023onebit,9930674}. 
The work in \cite{yang2023plugin} validates that the covariance recovery technique from \cite{Dirksen21} can be used in combination with the Bussgang estimator in order to perform channel estimation.
However, the considered quantizer designs with non-zero thresholds are challenging to implement in communication systems and may require more sophisticated analog and digital signal processing, ultimately resulting in performance losses. In contrast, considering multi-bit quantization of the input signal, coarse amplitude information is preserved because of the multi-level quantization, even with a fixed zero-threshold. However, up to now, there is no covariance recovery algorithm proposed for this case.

We derive a novel low-complexity covariance recovery algorithm for the multi-bit case by splitting the task into estimating the correlation matrix and the variances independently. 
Let us define the following problem statement where we reuse the notation from above for simplicity.
Consider a dataset $\mathcal{R} = \{\B r_t\}_{t=1}^T$ of $T$ samples of the form $\B r_t = Q_B(\B y_t)$ where $\B y_t\sim\NC(\B 0, \B C_{\B y})$ and $B > 1$. The task is to recover $\B C_{\B y}$ from the $T$ quantized samples.
Since no closed-form solution for the correlation matrix $\B R_{\B y} = \diag(\B C_{\B y})\sqrtm \B C_{\B y}\diag(\B C_{\B y})\sqrtm$
in the case of multi-bit quantization exists, we simply discard the samples' amplitude information, effectively treating them as one-bit quantization data. Because of that, the closed-form expression for estimating the unquantized correlation matrix $\B R_{\B y}$ by means of the one-bit sample covariance matrix $\hat{\B C}_{\text{1bit}} = \frac{1}{T}\sum_{t=1}^T Q_1(\B r_t) Q_1(\B r_t)\h$ can be obtained via the inverse arcsine law:
\begin{align}\label{eq:inv_arcsine}
    \hat{\B R}_{\B y} = \sin\left(\frac{\pi}{2} \Re(\hat{\B C}_{\text{1bit}})\right) + \op j \sin\left(\frac{\pi}{2} \Im(\hat{\B C}_{\text{1bit}})\right).
\end{align}
Unfortunately, although having a closed-form solution, the resulting correlation estimate is not necessarily \ac{psd} \cite{Dirksen21,PSD_spheres}. However, this can be resolved by a projection onto the set of \ac{psd} matrices, as discussed later.

Since the quantization acts elementwise on the real and imaginary part independently, it is sufficient to derive the variance estimation for a real-valued scalar $y\sim \mathcal{N}(0, \xi^2)$ and $r = Q_B(y)\in \mathcal{R}$ for ease of notation.
We note that the amplitude of $y$ follows the half-normal distribution.
The corresponding \ac{cdf} of the half-normal distribution is given by $\op P(|y|\leq \tau) = \erf(\tau / \sqrt{2\xi^2})$. 
Because the \ac{cdf} is fully parameterized by the input signal's variance $\xi^2$, one can utilize the coarse amplitude information after the quantizer for its estimation. 
By defining the positive quantization thresholds as $\tilde{\tau}_i<\infty$, $i\in\{1,\dots,2^{B-1}-1\}$, i.e., $\tilde{\tau}_i = \tau_{i + 2^{B-1}}$, one can estimate the probability of observing a sample with an amplitude of at most $\tilde{\tau}_i$ by 
$\hat{\op P}(|y| \leq \tilde{\tau}_i) = \frac{1}{T}\sum_{i=1}^T \chi(|r_t| \leq \tilde{\tau}_i)$.

For circularly symmetric Gaussian distributed complex-valued input and multiple quantization thresholds, i.e., $B>2$, an overdetermined system of equations can be constructed using the different quantizer thresholds for both the real- and imaginary parts, yielding $2^{B} - 2$ equations. 
The subtraction of two comes from the fact that the last quantization regions up to infinity are uninformative since, in this case, the (sample) probability is always one.
In summary, the equation system accounting for the real part of the input is of the form 
\begin{equation}
    \begin{aligned}
    \erf\left(\frac{\tilde{\tau}_i}{\sqrt{2\xi^2}}\right) = 
    \frac{1}{T}\sum_{t=1}^T\chi\left(|\Re(r_t)| \leq \tilde{\tau}_i\right)
    \end{aligned}
\end{equation}
with $i\in\{1,\dots,2^{B-1}-1\}$. The remaining half of the equation system is built similarly by replacing the real with the imaginary part.
Note that the equation system only depends on the unknown variance parameter $\xi^2$. Geometrically, we aim to interpolate the sample probabilities belonging to the different thresholds by a Gaussian \ac{cdf} curve in a \ac{ls} sense with the adjustable variance parameter $\xi^2$.
Since the derivative of the \ac{cdf} is trivially given by the Gaussian \ac{pdf}, a simple Gauss-Newton approach can be utilized for solving the nonlinear \ac{ls} problem.
For a robust initial starting point $\xi_0^2$, a solution to the equation with the quantizer's largest $\tilde{\tau}_i$ is used. 

In the multi-dimensional case, if the input signal's variance is assumed to be different for each antenna, the nonlinear \ac{ls} problem can be solved for each dimension independently, yielding an estimate of $\diag({\B C}_{\B y})$. Otherwise, the equation systems for each dimension can be combined to yield a more accurate estimate of the single variance parameter. 
Note that in the complex-valued case, the estimated variance has to be scaled by a factor of two to account for the sum of the real and imaginary parts.
Finally, the full covariance matrix estimate is computed as
\begin{equation}
	\hat{\B {C}}_{\B y} = \diag(\hat{\B {C}}_{\B y})^{\frac{1}{2}} \hat{\B R}_{\B y} \diag(\hat{\B {C}}_{\B y})^{\frac{1}{2}}.
\end{equation}

The derived covariance recovery scheme is now used in order to fit the \ac{gmm}'s covariances by only using quantized data. For simplicity, we assume that the training data stems from single snapshot observations, i.e., $\B r = Q_B(\B h + \B n)$ with $P=1$, which can always be enforced by pre-processing.
In each iteration of the \ac{em} algorithm, the M-step is adapted by using the proposed covariance recovery algorithm for estimating the unquantized covariance matrix due to the Gaussianity of each \ac{gmm} component. The necessary change to the purely Gaussian setting from before is that the responsibilities, computed in each E-step, are used in order to weight the sample probability for component $k$ accordingly as
\begin{equation}
    \hat{\op P}(|\Re([\B y]_n)| \leq \tilde{\tau}_i|k) = \frac{1}{N_k}\sum_{t=1}^T p(k|\B r_t) \chi(|\Re( [\B r_{t}]_n)| \leq \tilde{\tau}_i)
    \label{eq:samp_prob_gmm}
\end{equation}
for all $ i\in\{1,\dots,2^{B-1}-1\}$, where $N_k = \sum_{t=1}^T p(k|\B r_t)$. Note that $\Re([\B r_t]_n)$ is replaced by $\Im([\B r_t]_n)$ for the second half of the equation system.
Since the quantizer's input signal is also distorted with \ac{awgn}, the M-step adaptation from \cite[Th. 1]{10078293} for noisy data is used in addition by means of subtracting the noise covariance and afterward projecting to the set of \ac{psd} matrices by performing an \ac{evd} and truncating the negative eigenvalues. This also accounts for the possibly non-\ac{psd} correlation estimate from \eqref{eq:inv_arcsine}. 
After estimating the channel covariance $\hat{\B C}_{\B h|k}$ of component $k$ in this way, 
one first determines $\hat{\B C}_{\B y|k}$ to eventually construct the covariance of the quantized observation $\hat{\B C}_{\B r|k}$ by using one of the approximations given in \eqref{eq:quant_variance} or \eqref{eq:cr_multibit}.
Since the complexity is not crucial in the offline learning, we utilize the accurate formula for the variance \eqref{eq:quant_variance}.
The necessary adaptations in the M-step are concisely summarized in Algorithm~\ref{alg:mstep_quant}.
The so-found covariance matrix $\hat{\B C}_{\B r|k}$ is afterward used to compute the responsibilities in the E-step, similar to~\eqref{eq:resp_r}.

By the law of large numbers, the variance estimate based on the sample probability is a consistent estimator and, together with the closed-form solution for the correlation matrix, yields a consistent covariance estimator. This is a very convenient property since pilot observations are received in large numbers during regular operation at the \ac{bs}, and thus, a large training dataset $\mathcal{R}$ can be constructed.

\begin{algorithm}[t]
	\caption{Adapted M-step for quantized training data.}
	\label{alg:mstep_quant}
	\begin{algorithmic}[1]
		\REQUIRE $\mathcal{R} = \{\B r_t\}_{t=1}^T$, $\sigma^2$
            \vspace{0.05cm}
        \STATE Get $\{p(k|\B r_t)\}_{t=1,k=1}^{T,K}$ from E-step via \eqref{eq:resp_r}.
		\renewcommand{\algorithmicendfor}{\textbf{end}}
		\renewcommand{\algorithmicendwhile}{\textbf{end}}
		\FOR{$k=1$ to $K$}
		\STATE $N_k = \sum_{t=1}^T p(k|\B r_t)$
        \STATE $\hat{\B C}_{\B r|k}^{\text{1bit}} = \frac{1}{N_k}\sum_{t=1}^T p(k|\B r_t)Q_{1}(\B r_t)Q_1(\B r_t)\h$
        \STATE $\hat{\B R}_{\B y |k} = \sin\left(\frac{\pi}{2} \Re(\hat{\B C}_{\B r|k}^{\text{1bit}})\right) + \op j \sin\left(\frac{\pi}{2} \Im(\hat{\B C}_{\B r|k}^{\text{1bit}})\right)$
        \FOR{$n=1$ to $N$}
        \STATE Construct equation system for $\diag(\hat{\B C}_{\B y|k})$ via \eqref{eq:samp_prob_gmm}.
        \STATE Solve equation system via Gauss-Newton.
        \ENDFOR
        \STATE $\hat{\B {C}}_{\B y|k} = \diag(\hat{\B {C}}_{\B y|k})^{\frac{1}{2}} \hat{\B R}_{\B y|k} \diag(\hat{\B {C}}_{\B y|k})^{\frac{1}{2}}$
        \STATE $\B V\diag(\hat{\B c}_{\B h|k})\B V\h = \operatorname{EVD}(\hat{\B C}_{\B y|k} - \sigma^2\eye)$
        \STATE $\hat{\B c}_{\B h|k}^{\text{PSD}} = \max(\B 0, \hat{\B c}_{\B h|k})$ (elementwise max)
        \STATE $\hat{\B C}_{\B h|k} = \B V\diag(\hat{\B c}_{\B h|k}^{\text{PSD}})\B V\h$
        \STATE $\hat{\B C}_{\B y|k} = \hat{\B C}_{\B h|k} + \sigma^2 \eye$
        \STATE Compute $\diag(\hat{\B C}_{\B r|k})$ from $\hat{\B C}_{\B y|k}$ via \eqref{eq:quant_variance}.
        \STATE Compute $\B B_{k}$ from $\hat{\B C}_{\B y|k}$ via \eqref{eq:Buss_multi}.
        \STATE $\nondiag(\hat{\B C}_{\B r|k})= \nondiag(\B B_k \hat{\B C}_{\B y|k} \B B_k\h)$
        \ENDFOR
	\end{algorithmic}
\end{algorithm}

\subsection{Loss Function Adaptation for the VAE}\label{subsec:vae_real}
Similar to the \ac{gmm}, the \ac{vae} model can be trained directly from noisy pilot observations by properly adapting the \ac{elbo} loss function \cite{baur2023leveraging,10051858}. The idea, thereby, is to modify the parameterized channel's covariance matrix, which stems from the output of the decoder such that it represents the covariance matrix of the pilot observations
in the loss function \eqref{eq:elbo}. 
Consequently, only the channel covariance matrix is learned by the \ac{vae} through gradient updates. For the case of coarsely quantized training samples from $\mathcal{R}$, the expression of the quantized covariance matrix with respect to the channel covariance matrix is not in closed-form and thus not differentiable, cf. \eqref{eq:quant_variance}. However, one can use the approximation from \eqref{eq:cr_multibit} as shown in the following. 

Since the training is done in the Fourier-transformed domain in order to parameterize a circulant covariance matrix $\B C_\theta  =\dft\h \diag(\B c_\theta) \dft$, we first simplify the expression for the diagonal term $\diag(\B C_\theta) = \diag(\dft\h \diag(\B c_\theta) \dft) = \frac{1}{N}\sum_{n=1}^N [\B c_\theta]_n\eye$, which can be simply shown by writing $\diag(\B C_\theta) = \frac{1}{N}\tr(\B C_\theta)\eye$ and utilizing the properties of the trace together with the unitary \ac{dft} matrix $\dft$.
Using this, we can now write $\diag(\B C_{\B y,\theta}) = \frac{1}{N} \sum_{n=1}^N ([\B c_{\theta}]_n + \sigma^2)\eye$ when assuming single snapshot pilot observations as $\B r = Q_B(\B h + \B n)$, such that $\B C_{\B r,\theta}=\dft\h \diag(\B c_{\B r,\theta})\dft$ where $\B c_{\B r,\theta} $ equals
\begin{equation}\label{eq:vae_cr}
    (1-\rho_\theta)^2(\B c_\theta + \sigma^2\B 1) + \rho_\theta (1 - \rho_\theta) \frac{1}{N} \sum_{n=1}^N ([\B c_{\theta}]_n + \sigma^2)\B 1
\end{equation}
and where $\rho_\theta$ is computed via \eqref{eq:Buss_multi} by plugging in $\diag(\B C_{\B y,\theta})$. Thus, the \ac{vae} model can be learned from quantized data by replacing $\B c_\theta$ with $\B c_{\B r,\theta}$ from \eqref{eq:vae_cr} and $\B h$ with $\B r$ in the loss function \eqref{eq:elbo}.

\section{Baseline Channel Estimators}
\label{sec:baselines}

We compare the proposed parameterized generative modeling-aided channel estimators with state-of-the-art baseline channel estimators for coarse quantization systems. 
    First, for the 3GPP channel model from \Cref{sec:channel_model}, we have genie access to the true underlying channel covariance matrix $\B C_{\B h|\B \delta}$ from \eqref{eq:3gpp_cov} in the simulation. This allows us to evaluate the genie-aided Bussgang estimator $\hat{\B h}_{\text{Buss-genie}} = \B C_{\B h| \B \delta} \B A\h \B B_{\B \delta}\h \B C_{\B r|\B \delta}\inv \B r$
    where $\B B_{\B \delta}$ and $\B C_{\B r|\B \delta}$ are found by plugging $\B C_{\B h|\B \delta}$ into \eqref{eq:Buss_multi} or \eqref{eq:Buss_one} and \eqref{eq:cr_multibit} or \eqref{eq:arcsin_law} for the multi-bit or one-bit case, respectively. Note that this estimator is not feasible in practice but only serves as a lower bound on the performance of the Bussgang estimator, which is the best linear estimator. The corresponding curves are labeled as ``Buss-genie''.
    
    A practicably feasible approach that is primarily used in the literature is to use the sample covariance matrix $\hat{\B C}_{\B h} = \frac{1}{T}\sum_{t=1}^T \B h_t\B h_t\h$ in combination with the Bussgang estimator \eqref{eq:h_buss}, labeled as ``Buss-Scov''. Note that for this case, a training dataset of ground-truth channels $\mathcal{H}$ is necessary. 
    
    A simple baseline is the \ac{ls} estimate based on the Bussgang decomposition \eqref{eq:Bussgang_decomp}, i.e., $\hat{\B h}_{\text{BLS}} = \B A^\dagger \B B^\dagger \B r$, labeled as ``BLS''. For computing the Bussgang gain \eqref{eq:Buss_multi} or \eqref{eq:Buss_one}, we use the sample covariance matrix $\hat{\B C}_{\B h}$ from above. 

    In \cite{Mo2018}, a \ac{cs}-based channel estimator is proposed, which is a combination of the \ac{em} algorithm for approximating the channel \ac{pdf} in the sparse angular domain and the \ac{gamp} algorithm to solve the sparse recovery problem. 
    Note that in this case, an EM algorithm is deployed online for each transmission link and pilot observation, which is fundamentally different from the \ac{gmm} approach, which exploits the \ac{em} solely in the offline phase.
    We applied the EM-GM-GAMP algorithm to estimate the channel parameters $\hat{\B x}$ in the angular domain, such that the final channel estimate is computed to $\hat{\B h}_{\text{EM-GM-GAMP}} = \dft \hat{\B x}$, labeled as ``EM-GM-GAMP''.
    
    We also evaluate a deep learning-based estimator, similar to \cite{Zhang2020}, where a three-layered feed-forward \ac{nn} is trained to directly map the pilot observation to a channel estimate. The \ac{relu} function is used as the activation function in all layers except the output layer. To achieve a fair comparison to the proposed approaches, a single network is trained for the whole \ac{snr} range. Similar to \cite{Zhang2020}, the best performance was achieved by a drastic increase of the neurons in the hidden layers. We, therefore, set the number of neurons in both hidden layers to $2N^2$. The corresponding curves are labeled as ``DNN''.

\section{Memory and Complexity Analysis}
\label{sec:memory_complexity}

\begin{table*}[t]
	\centering
	\renewcommand{\arraystretch}{1.1}
	  	\caption{Computational complexity and number of parameters of the discussed estimators with example numbers for the case of $K=N=64$, $L=16$, and $P=4$.}  
    \resizebox{1\textwidth}{!}{
                \begin{tabularx}{1\textwidth}{|X|m{0.38\textwidth}|X|m{0.25\textwidth}|}
	    \hline 
		\textbf{Name} & \textbf{Model Parameters (real-valued)} & \textbf{Example (rounded)} &\textbf{Online Complexity}
		\\
		\hline 
		GMM full &$K\left(N^2+1\right)-1$ & $2.62\cdot 10^5$ &$\mathcal{O}(KPN^2)$ (parallel in $K$)
		\\
		\hline
		GMM toep &$K\left(4N + 1\right)- 1$ & $1.64\cdot 10^4$ &$\mathcal{O}(KPN^2)$ (parallel in $K$)
		\\
		\hline
		GMM circ &$K\left(N + 1\right)-1$ & $4.16\cdot 10^3$ &$\mathcal{O}(KPN\log (PN))$ (parallel in $K$)
		\\
		\hline
	    MFA &$K\left(2LN+2\right)-1$ & $1.31\cdot 10^5$ &$\mathcal{O}(KPN^2)$ (parallel in $K$)
		\\
		\hline
	    VAE & $6N^2 + 10NL + \frac{1}{2}P^2 + \frac{1}{2}P + \mathcal{O}(N)$
	    & $3.53\cdot 10^4$ &$\mathcal{O}(N^2 + PN\log(PN))$
		\\
		\Xhline{4\arrayrulewidth}
	    DNN \cite{Zhang2020}& $4N^4 + (4P + 4)N^3 + 4N^2 + 2N$ & $7.24\cdot 10^7$ &$\mathcal{O}( N^3)$
		\\
		\hline
	    Buss-Scov & $N^2$ & $5.0\cdot 10^{3}$ &$\mathcal{O}(PN^2)$
		\\
		\hline
	    EM-GM-GAMP \cite{Mo2018}& - & - &$\mathcal{O}(PN\log (PN))$
		\\
		\hline
	    BLS & - & - &$\mathcal{O}(PN)$
		\\
		\hline
	\end{tabularx}
	}
	\label{tab:memory}
\end{table*}

The offline memory requirements of data-based techniques and the algorithmic online complexity are key features for channel estimation in real-time systems. 
The number of parameters of the (structured) zero-mean \ac{gmm} and the \ac{mfa} model is determined by the $K$ covariances and the number of mixing coefficients. 
The corresponding linear \ac{mmse} filters for each component and \ac{snr} value are fixed after the offline training, which means that they can be pre-computed. Since this can be similarly done for the evaluation of the responsibilities in \eqref{eq:resp_r}, the overall online complexity is determined by matrix-vector products for each component~\cite{9842343}. Notably, the computation of the $K$ filters/responsibilities is trivially parallelizable, which is of great importance in practical systems.
For the case of circulant-structured \ac{gmm} covariances, cf. \Cref{subsec:gmm}, the complexity reduces due to the usage of \acp{fft} \cite{9842343}. 

For the \ac{vae} approach, the number of parameters and the complexity for a forward pass through the network depend on the network architecture, cf. \Cref{subsec:vae}. The resulting filter is computable by means of \acp{fft} since a circulant covariance matrix is parameterized, similar to the circulant \ac{gmm} case. We further note that the approaches that learn from quantized data, cf. \Cref{sec:learning_quant}, are only adapted in the training procedure and thus have the same memory overhead and online complexity as the models learned with perfect \ac{csi}.

Table \ref{tab:memory} summarizes the memory overhead and computational online complexity of all proposed approaches as well as the baseline methods. It can be seen that the proposed approaches vary in the number of parameters and online complexity to allow for a smooth trade-off with respect to the desirable performance and practical system requirements. Of particular importance is the comparison to the deep \ac{nn} approach, which is adapted from \cite{Zhang2020} and directly provides a channel estimate at the output, i.e., it does not parameterize an analytical estimator. It becomes apparent that the proposed approaches exhibit a much lower number of parameters as well as a lower online complexity compared to the deep \ac{nn} approach. The main reason for this is the drastic increase of neurons in the hidden layers in the \ac{nn} approach, cf. \cite{Zhang2020}; in contrast, the proposed models are comprised of a latent space which enforces a compression, and thus, a reduced memory and complexity overhead. As shown in the following numerical results, the estimation performance of the proposed approaches is, in most cases, even better, although having fewer parameters and reduced online complexity.

\section{Achievable Rate Lower Bound}
The achievable rate is of great interest in quantized systems~\cite{Li2017,7894211}. 
We evaluate a lower bound on the corresponding achievable rate of a respective data transmission system that is taking the \ac{csi} mismatch into account. To this end, after estimating the channel with the pilot transmission in \eqref{eq:sytem_model}, the data symbol $s$ is transmitted over the same channel, i.e.,
$\B r = Q_B(\B h s + \B n) =  \B B \B h s + \B q$; in the second equation, the linearized model with Bussgang's decomposition is used where $\B q = \B B \B n + \B \eta$. We make the worst-case assumption that the aggregated noise is Gaussian, i.e., $\B q \sim \mathcal{N}_\mathbb{C}(\B 0, \B C_{\B q} = \B C_{\B r}-\B B \B C_{\B h} \B B\h )$, cf. \cite{1193803}.
Furthermore, the \ac{bs} is assumed to perform \ac{mrc} with the normalized filter $\B g_{\text{MRC}}\h = \hat{\B h}\h / \|\hat{\B h}\|_2^2$.
Note that the variance of the data symbol $s$ is assumed to be one without loss of generality. We further assume that the \ac{snr} is the same during pilot and data transmission.
Thus, we can evaluate the \ac{uatf} bound as a lower bound on the achievable rate, cf. \cite[Lemma 1]{Li2017}, as
\begin{equation}
    R_{\text{UF}} = \log_2\left(1 + \frac{\big|\op E [\B g_{\text{MRC}}\h \B B \B h]\big|^2}{
    \operatorname{var}[\B g_{\text{MRC}}\h \B B \B h] + \op E[\B g_{\text{MRC}}\h \B C_{\B q} \B g_{\text{MRC}}]}\right)
    \label{eq:rate}
\end{equation}
with Monte Carlo simulations.

\section{Simulation Results}\label{sec:sim_results}

\subsection{Covariance Recovery}\label{subsec:sim_results_cov}

\begin{figure}[t]
	\centering
    \includegraphics{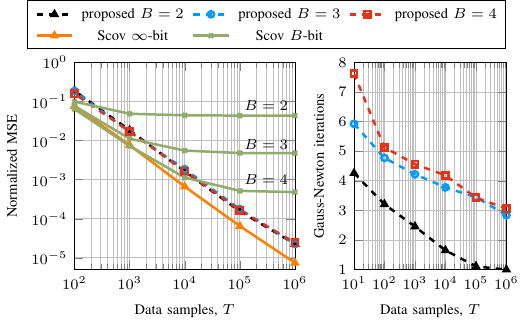}
	\caption{Performance evaluation for covariance estimation with $N=64$ dimensions and $100$ Monte Carlo iterations using covariances obtained from the 3GPP model \eqref{eq:3gpp_cov}.}
	\label{fig:cov_recovery_mse_iter}
\end{figure}

\begin{figure*}[t]
    \centering
    \includegraphics[width=\textwidth]{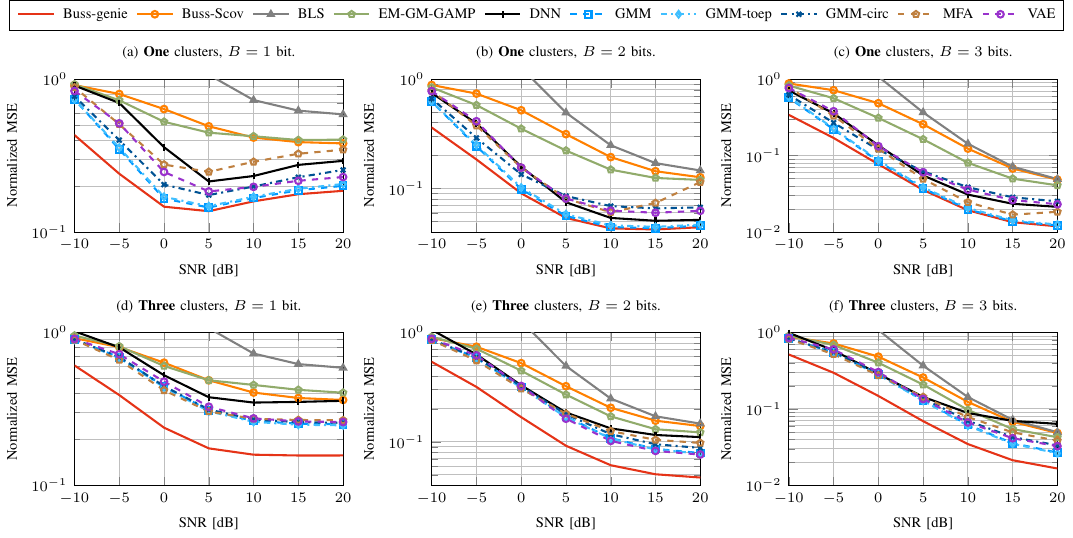}
	\caption{\ac{mse} performance for the 3GPP channel model (cf. \Cref{sec:channel_model}) with one (top) and three (bottom) propagation clusters for $B\in\{1,2,3\}$ quantization bits, $N=64$ antennas, and $P=1$ pilot observation (single snapshot).}
	\label{fig:mse_bits_3gpp}
\end{figure*}

\begin{figure}[t]
    \centering
    \includegraphics{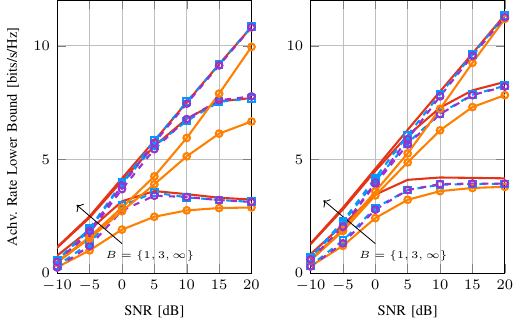}
    \caption{Evaluation of the achievable rate lower bound \eqref{eq:rate} for the 3GPP channel model (cf. \Cref{sec:channel_model}) with one (left) and three (right) propagation clusters for $B=\{1,3,\infty\}$ quantization bits, $N=64$ antennas, and $P=1$ pilot observation.}
	\label{fig:rate_3gpp}
\end{figure}

\begin{figure}[t]
    \centering
    \includegraphics[width=1\columnwidth]{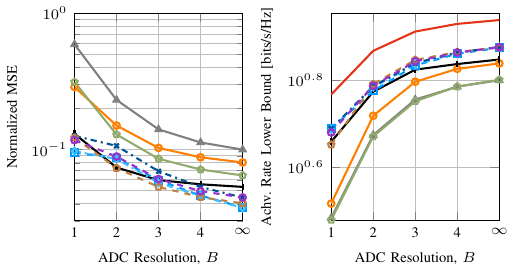}
	\caption{\ac{mse} (left) and achievable rate lower bound (right) performance for the QuaDRiGa LOS channel model (cf. \Cref{sec:channel_model}) for $N=64$ antennas and $P=1$ pilot.}
	\label{fig:adcs_quadriga}
\end{figure}

Before investigating the channel estimation performance, we evaluate the proposed covariance recovery algorithm from \Cref{subsec:gmm_quant} in a purely Gaussian setting without \ac{awgn}, comparing it to reasonable baselines. 
By assuming genie-knowledge of the unquantized samples, we can evaluate the unquantized sample covariance matrix, i.e., $\hat{\B C}_{\text{unquant}} = \frac{1}{T}\sum_{t=1}^T \B h_t\B h_t\h$. Note that this baseline requires perfect \ac{csi} and thus only serves as a baseline. 
A feasible approach is to neglect the quantization effect and evaluate the quantized sample covariance matrix $\hat{\B C}_{\text{quant}} = \frac{1}{T}\sum_{t=1}^T \B r_t\B r_t\h$, where $\B r_t = Q_B(\B h_t)$. This baseline becomes more accurate with more quantization bits $B$ but introduces a systematic error due to the coarse quantization.

We construct $100$ random covariance matrices $\B C_{\B h |\B \delta}$ from~\eqref{eq:3gpp_cov} and draw a fixed number of samples from each covariance matrix as $\{\B h_t \sim \NC(\B 0, \B C_{\B h|\B \delta})\}_{t=1}^T$. Afterward, the normalized \ac{mse} $\op E[\|\B C_{\B h|\B \delta} - \hat{\B C}\|_F^2] / \op E [\|\B C_{\B h|\B \delta}\|_F^2]$ is computed by using those $100$ covariance realizations.

The left plot in Fig \ref{fig:cov_recovery_mse_iter} shows the normalized \ac{mse} versus different numbers $T$ of samples for different numbers $B$ of quantization bits. It can be seen that the proposed covariance recovery scheme performs equally well for all quantization levels since it yields a consistent estimator, i.e., the estimation error steadily decreases for a larger number of samples. 
This is similar to the unquantized sample covariance matrix (``Scov $\infty$-bit'') but with a more or less constant offset, which is mainly caused by the correlation estimate that is unchanged for varying bits $B$, cf. \eqref{eq:inv_arcsine}.
In contrast, the quantized sample covariance matrix (``Scov $B$-bit'') is a biased estimator and shows a relatively high error floor, which decreases for a higher number of quantization bits, as excepted. 
We note that the consistency and unbiasedness of the proposed estimator is a key characteristic since, for the training with quantized pilot observations, it can be expected that a large dataset $\mathcal{R}$ can be acquired cheaply during regular operation of the \ac{bs}; this is in contrast to a dataset $\mathcal{H}$ consisting of ground-truth channels, which either requires costly measurement campaigns or intricate modeling of the underlying propagation environment.

In the right plot in Fig. \ref{fig:cov_recovery_mse_iter}, we evaluate the necessary number of iterations of the Gauss-Newton algorithm for solving the nonlinear \ac{ls} problem until convergence, i.e., until the absolute change of the estimated variances is smaller than $10^{-5}$. 
It can be seen that in all cases, only a few iterations are necessary for the convergence; the number of iterations also decreases for a higher number of data samples $T$ and for less quantization bits (due to an increasing number of equations for larger $B$), which makes the variance estimation very fast. 
In combination with the closed-form solution for the correlation matrix \eqref{eq:inv_arcsine}, the covariance estimator exhibits low complexity and is applicable for any given number $B>2$ of quantization bits.

\subsection{Channel Estimation}\label{subsec:sim_resuts_chEst}

This section provides numerical results to evaluate the proposed channel estimators, cf. \Cref{sec:main_part} and \Cref{sec:learning_quant}, against the discussed state-oft-the-art baselines from \Cref{sec:baselines}. In all simulations, we have fixed the number of training samples for both $\mathcal{H}$ and $\mathcal{R}$, cf. \Cref{subsec:datasets}, to $T=100{,}000$. 
The normalized \ac{mse} $\frac{1}{NT_{\text{test}}} \sum_{i=1}^{T_{\text{test}}} \|\B h_i - \hat{\B h}_i\|_2^2$ 
and the achievable rate lower bound \eqref{eq:rate} are computed by means of $T_{\text{test}} = 10{,}000$ channel samples, which are not part of the training dataset.
If not otherwise stated, the number of components for the \ac{gmm}/\ac{mfa} is $K=64$, the latent dimension for the \ac{mfa}/\ac{vae} is $L=N/4$, and the data-aided approaches are trained with the channel dataset $\mathcal{H}$. 
For both the \ac{vae} and the DNN approach, a single \ac{nn} architecture is trained for the whole \ac{snr} range of $[-10, 20]$dB. 
Since a low pilot overhead is considered to be a key aspect in practical systems \cite{8337813}, we, therefore, especially focus on the single snapshot scenario in this paper.
The simulation code of the proposed channel estimators is publicly available.\footnote{\href{https://github.com/benediktfesl/Quantized_Channel_Estimation}{https://github.com/benediktfesl/Quantized\_Channel\_Estimation}}

In Fig. \ref{fig:mse_bits_3gpp}, we evaluate the \ac{mse} performance of the proposed channel estimators in comparison to the baseline methods over the \ac{snr} for the 3GPP channel model from \Cref{sec:channel_model} with one (top row) and three (bottom row) propagation clusters for $B\in \{1,2,3\}$ quantization bits, $N=64$ \ac{bs} antennas, and $P=1$ pilot. 
In all cases, the approaches ``BLS'', ``Buss-Scov'', and ``EM-GM-GAMP'' are outperformed with a considerable performance gap over the whole \ac{snr} range by the proposed approaches. 
This is due to the fact that the Bussgang theorem does not hold for the non-Gaussian distributed channels, which is assumed by ``Buss-Scov''; moreover, the channels are generally not perfectly sparse in the angular domain (leakage effect), which substantially impacts the \ac{cs} approach ``EM-GM-GAMP''. Interestingly, for the case of one propagation cluster, the \ac{gmm}-based approach is close to the ``Buss-genie'' approach,
which is the Bussgang estimator with utopian knowledge of the true channel covariance matrix for a single snapshot; this underlines the powerful estimation abilities of the \ac{gmm}. For the considered case of a \ac{ula}, the Toeplitz-structured \ac{gmm} version is almost on par with the full \ac{gmm} approach, whereas the circulant-structured approach exhibits a small performance gap. 
The observation of having increasing \ac{mse} for higher \ac{snr} values beyond a certain \ac{snr} level in some cases is due to stochastic resonance, which is a well-known effect in quantized systems \cite{mcdonnell_stocks_pearce_abbott_2008}; thereby, the effect can vary for different estimators, depending on the parameterization.  
The ``DNN'' approach also shows good estimation results for the different scenarios but is consistently outperformed by at least one of the proposed estimators, although having a much larger number of parameters and a higher online complexity, cf. Table \ref{tab:memory}.
Overall, the simulation results in Fig.~\ref{fig:mse_bits_3gpp} demonstrate the great potential of the proposed class of parameterized estimators based on Gaussian latent models in combination with the Bussgang estimator. 

Fig. \ref{fig:rate_3gpp} assesses the achievable rate lower bound from \eqref{eq:rate} for the 3GPP channel model with one (left) and three (right) propagation clusters for $N=64$ antennas, $P=1$ pilot observation, and different numbers of quantization bits. For the case of one propagation cluster, the achievable rate lower bound of the \ac{gmm} and \ac{vae} approach is almost on par with the ``Buss-genie'' approach. Moreover, a substantial gap to the achievable rate lower bound of the ``Buss-Scov'' approach is apparent for all considered numbers of quantization bits. 
This behavior similarly translates to the case of three propagation clusters but with an overall reduced gap to the baseline approach. These results indicate that the better estimation performance of the proposed estimators can be effectively converted a higher data rate or to a lower resolution while preserving the same throughput as the baseline approach ``Buss-Scov''.

\begin{figure}[t]
    \centering
    \includegraphics[width=1\columnwidth]{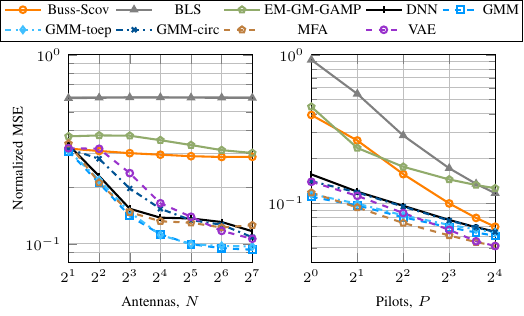}
		\caption{\ac{mse} performance for the QuaDRiGa LOS channel model (cf. \Cref{sec:channel_model}) for $B=1$. Left: $P=1$ and $\text{SNR} = 10$dB; right: $N=64$ and $\text{SNR} = 5$dB.}
	\label{fig:mse_antennas_pilots}
\end{figure}

\begin{figure}[t]
    \centering
    \includegraphics[width=1\columnwidth]{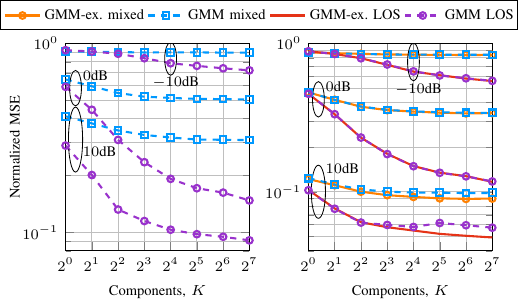}
	\caption{\ac{mse} performance for the QuaDRiGa LOS and mixed LOS/NLOS channel model (cf. \Cref{sec:channel_model}) for $B=1$ (left) and $B=3$ (right), $N=64$, and $P=1$.}
	\label{fig:components}
\end{figure}

\begin{figure*}[t]
    \centering
    \includegraphics[width=1\textwidth]{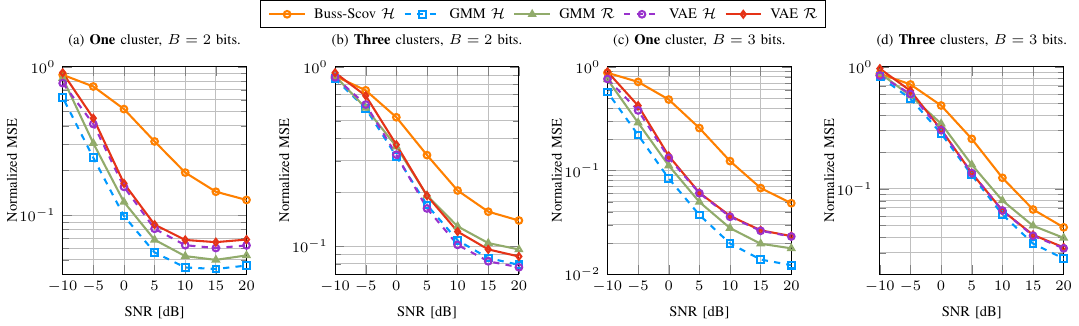}
	\caption{\ac{mse} performance evaluation of the models trained with quantized data for the 3GPP channel model (cf. \Cref{sec:channel_model}) for $P=1$ pilot and $N=64$ antennas.}
	\label{fig:mse_quant_data}
\end{figure*}

In Fig. \ref{fig:adcs_quadriga} (left), the \ac{mse} performance is compared for different numbers $B$ of quantization bits, now for the QuaDRiGa \ac{los} channel model, cf. \Cref{sec:channel_model}.
Once again, the approaches ``BLS'', ``Buss-Scov'', and ``EM-GM-GAMP'' are outperformed over the whole range of quantization bits. Interestingly, the ``DNN'' approach is comparably good for $B=2$ and $B=3$ but, in turn, suffers in performance in the extreme cases of $B=1$ and infinite resolution, which indicates better robustness of the proposed approaches in comparison. Interestingly, the \ac{mfa} estimator is ranked among the best estimators in this case, which highlights the fact that estimators' performances may vary slightly for different channel models; however, it can also be seen that the overall performance of the proposed class of estimators is stable and robust with respect to a different channel model.
In Fig. \ref{fig:adcs_quadriga} (right), the corresponding achievable rate lower bound from~\eqref{eq:rate} is evaluated. It can be seen that the better estimation qualities of the proposed approaches in terms of the \ac{mse} directly translate to a higher achievable rate guarantee, which is approximately only 1--2 bits/s/Hz below the achievable rate lower bound with perfect \ac{csi} knowledge at the receiver.

The left plot in Fig. \ref{fig:mse_antennas_pilots} examines the estimation quality over different numbers $N$ of antennas at the \ac{bs} for $B=1$ bit, $P=1$ pilot, and an \ac{snr} of $10$dB for the QuaDRiGa \ac{los} channel model, cf. \Cref{sec:channel_model}. In contrast to the baseline approaches ``BLS'', ``Buss-Scov'', and ``EM-GM-GAMP'', the performance of the proposed estimators significantly increases for a higher number of antennas, which is particularly important in massive \ac{mimo} systems. 
The \ac{gmm} approach performs best for all considered antenna numbers, whereas the \ac{vae} approach is especially strong in the high number of antennas case; this is reasoned by the circulant parameterization of the covariances by the \ac{vae}, which only holds asymptotically for high numbers of antennas. 
A similar behavior is observed for the ``GMM-circ'' estimator. The ``DNN'' approach, which has a quartic scaling of the number of parameters in the number of antennas, cf. Table \ref{tab:memory}, is outperformed over the whole range.

The right plot in Fig. \ref{fig:mse_antennas_pilots} shows the \ac{mse} performance for an increasing number of pilot observations by utilizing the pilot design from \Cref{subsec:system_model} for $N=64$ antennas and a fixed \ac{snr} of $5$dB. The proposed estimators outperform all baseline approaches over the whole range of pilot observations, including the ``DNN'' estimator. Especially the \ac{mfa} and \ac{vae} models, which are comprised of a nonlinear latent space, perform well in the high number of pilots regime.
Although we focus our analysis primarily on the single-snapshot case, we see that also with an increasing number of pilots, the proposed estimators perform very well.

Next, the number $K$ of \ac{gmm} components that are necessary to achieve a certain performance is discussed. In the left plot of Fig. \ref{fig:components}, the \ac{mse} over the number of \ac{gmm} components is evaluated for $B=1$ bit, $N=64$ antennas, $P=1$ pilot, and for varying \acp{snr} for the QuaDRiGa \ac{los} as well as mixed \ac{los}/\ac{nlos} channel model, cf. \Cref{sec:channel_model}. 
It can be expected that the superposition of many sub-paths, as it is the case in the mixed \ac{los}/\ac{nlos} scenario, results in a less structured wireless channel, and thus, less structural information can be inferred as prior knowledge by the data-aided models. 
Therefore, it can be observed that the overall performance is worse for the mixed scenario. However, in both scenarios, the increase of \ac{gmm} components continuously enhances the estimation performance, with a greater improvement in the pure \ac{los} case. This points towards applications in \ac{mmwave} communications, where the high frequency in combination with smaller \ac{bs} cells results in high \ac{los} probabilities.

The right plot of Fig. \ref{fig:components} analyzes the same setup but now for $B=3$ quantization bits. In this case, we evaluate the approximation quality of computing $\B C_{\B r}$ for a given $\B C_{\B y}$ via \eqref{eq:cr_multibit} by comparing it with the estimator that computes the exact variances via \eqref{eq:quant_variance} and otherwise uses the same approximation for the off-diagonals, labeled ``GMM-ex.''. 
As expected, the approximation is highly accurate in the low to medium \ac{snr} region, which is the considered operating range of low-resolution systems. In high \ac{snr}, the approximation is less accurate and shows saturation effects when increasing the number of \ac{gmm} components. 
However, the overall approximation loss is small, and it still results in a high estimation accuracy as compared to the baseline approaches. Besides that, the estimation performance generally steadily increases for a higher number of \ac{gmm} components with an overall saturation for high numbers $K$ of components.

Fig. \ref{fig:mse_quant_data} evaluates the proposed training adaptations for the \ac{gmm} and the \ac{vae}, as detailed in \Cref{sec:learning_quant}, in order to learn from noisy and quantized pilot observations $\mathcal{R}$, cf. \Cref{subsec:datasets}, without having ground-truth channel samples during training. 
We refer to the adapted \ac{gmm}, cf. Algorithm~\ref{alg:mstep_quant}, as ``GMM $\mathcal{R}$'', and the \ac{gmm} learned with ground-truth channels as ``GMM $\mathcal{H}$''. The \acp{vae} are denoted likewise.
Note that a main difference between ``GMM $\mathcal{R}$'' and ``VAE $\mathcal{R}$ is that the training of the adapted \ac{gmm} is performed for a fixed \ac{snr}, but afterward, the model can be utilized for the whole \ac{snr} range, whereas the \ac{vae} is trained directly for the whole \ac{snr} range in order to generalize properly. In order to have a meaningful comparison, the ``GMM $\mathcal{R}$'' is trained for each \ac{snr} point, whereas the ``VAE $\mathcal{R}$'' is trained for the whole \ac{snr} range (no performance gain was seen in the simulation results for an \ac{snr}-dependent training).

In Fig. \ref{fig:mse_quant_data} (a) and (b), the case of $N=64$, $P=1$, $B=2$, and the 3GPP channel model with one and three propagation clusters are considered, respectively, whereas in Fig. \ref{fig:mse_quant_data} (c) and (d) the same setup with $B=3$ bits is investigated. 
Astonishingly, through the model-based adaptations, the 
models trained with the dataset $\mathcal{R}$ consisting of coarsely quantized and noisy data samples are almost on par with their counterparts that are trained with ground-truth noise-free channel samples from $\mathcal{H}$.
Overall, the estimation quality seems to be most accurate in the low to medium \ac{snr} range, where the approximation in \eqref{eq:cr_multibit} and thus the training adaptations are highly accurate.
This implies that the generally costly dataset $\mathcal{H}$ can be replaced by $\mathcal{R}$ with almost no performance loss in this regime. In the high \ac{snr} regime, the performance loss of ``GMM $\mathcal{R}$'' tends to increase, which is a consequence of the generally indefinite closed-form correlation estimate \eqref{eq:inv_arcsine}. 
After the projection onto the set of \ac{psd} matrices by truncating the negative eigenvalues, cf. Algorithm \ref{alg:mstep_quant}, the resulting covariance estimate is missing the corresponding eigenvectors, which has a higher impact on the performance in the high \ac{snr} regime. 
This correlates with the observation that the adaptations are working best in cases with fewer multi-path components.

\section{Conclusion}\label{sec:conclusion}
In this work, we presented a novel and promising framework for channel estimation in coarse quantization systems by utilizing Gaussian latent models such as \acp{gmm}, \acp{mfa}, and \acp{vae}.
These models successfully learn the unknown and complex channel distributions present in radio propagation scenarios and, afterward, utilize this valuable prior information to enable the development of tractable parameterized linear \ac{mmse} estimators based on a conditional Bussgang decomposition.
We have shown that all of the presented estimators perform well for various channel and system parameters with only minor differences. This allows for selecting the preferred model using the discussed memory and complexity overhead.

In addition, we derived model-based training adaptations, i.e., a covariance recovery algorithm for the \ac{gmm} and a loss function adaptation for the \ac{vae}, in order to learn these models directly from quantized training data, with only marginal performance losses. 
Extensive simulations verified a superior performance over classical and deep learning-based approaches in terms of \ac{mse} and achievable rate metrics.

The presented work outlines several directions for further investigating the proposed estimation framework, e.g., the analysis of multi-user systems, pilot contamination in multi-cell systems, and different receive strategies, e.g., zero-forcing. With the recent advances of the discussed \ac{vae} concept to model time-varying channels \cite{10094953}, an extension of the presented estimation framework to time-varying systems is part of future work.

\section{Acknowledgment}
The authors gratefully acknowledge valuable discussions with and input from Dr.-Ing. Michael Koller in the early stages of this work.

\appendix

\subsection{Proof of Lemma \ref{lemma:cond_Buss}}\label{app:proof_lemma1}
\begin{proof}
Since $\B h|\B c$ is zero-mean Gaussian and $\B n$ is independent of $\B c$ by assumption, it follows that $\B y | \B c \sim \mathcal{N}_\C(\B y; \B 0, \B C_{\B y| \B c})$. Therefore, given $\B c$, the requirements for applying Bussgang's theorem are fulfilled and we can choose $\B B_{\B c} = \B C_{\B r\B y|\B c} \B C_{\B y|\B c}\inv$ such that $\B y$ and $\B \eta$ are conditionally uncorrelated given $\B c$. In particular, the conditional Bussgang gain $\B B_{\B c}$ can be computed via \eqref{eq:Buss_multi} or \eqref{eq:Buss_one}, respectively, and the closed-form expressions as well as the approximation in \eqref{eq:arcsin_law}--\eqref{eq:cr_multibit} hold for computing $\B C_{\B r|\B c}$ by substituting $\B C_{\B y}$ with $\B C_{\B y|\B c}$ because of the Gaussianity of the input given $\B c$, proving the properties a), b), and c).

We further note that since $\op E[\B y | \B c] = \B 0$ and the uniform quantization function $Q_B(\cdot)$ is symmetric around zero, it follows that $\op E[\B r | \B c] = \B 0$. Thus, the conditional linear \ac{mmse} estimator of $\B h$ given $\B r$ and $\B c$ is computed as 
\begin{align}
    \hat{\B h}_{\B c} &= \op E[\B h \B r\h |\B c] \op E[\B r \B r\h |\B c]\inv \B r
    \\
    &= \op E[\B h (\B h\h \B A\h\B B_{\B c}\h +\B n\h \B B_{\B c}\h +\B \eta\h ) |\B c] \B C_{\B r|\B c}\inv \B r
    \\
    &= (\B C_{\B h|\B c} \B A\h\B B_{\B c}\h + \op E[\B h \B \eta\h | \B c]) \B C_{\B r|\B c}\inv\B r,
    \label{eq:linearMMSE_last}
\end{align}
where \eqref{eq:linearMMSE_last} follows from the fact that $\op E[\B h\B n\h |\B c] = \op E[\B h| \B c] \op E[\B n|\B c]\h = \B 0$. Similar as in \cite[Appendix A]{Li2017} for the unconditional case, the conditional correlation of $\B h$ and $\B \eta$ given $\B c$ computes to
\begin{align}
    \op E[\B h \B \eta\h | \B c] &= \op E [\op E[\B h \B \eta\h | \B c, \B y] |\B c]
    \label{eq:corr_eta1}
    \\
    &= \op E [\op E[\B h | \B c, \B y]\B \eta\h  |\B c]
    \label{eq:corr_eta2}
    \\
    &= \B C_{\B h |\B c}\B A\h\B C_{\B y| \B c}\inv \op E[\B y\B \eta\h |\B c] = \B 0,
    \label{eq:corr_eta3}
\end{align}
where in \eqref{eq:corr_eta1} we use the law of total expectation, \eqref{eq:corr_eta2} holds since $\B \eta = \B r - \B B_{\B c}\B y$ is constant for a given $\B y$, and in \eqref{eq:corr_eta3} we use the fact that $\B h$ and $\B y$ are conditionally jointly Gaussian given $\B c$ and, thus, the conditional expectation $\op E[\B h | \B c, \B y]$ is computed by the conditional linear \ac{mmse} estimator. Since $\B y$ and $\B \eta$ are conditionally uncorrelated given $\B c$ by means of the Bussgang decomposition, the estimator in \eqref{eq:linearMMSE_last} simplifies to \eqref{eq:cond_lin_MMSE}, finishing the proof.
\end{proof}

\bibliographystyle{IEEEtran}
\bibliography{IEEEabrv,library}

\end{document}